\documentclass[pra,twocolumn,floatfix,a4paper,superscriptaddress]{revtex4}
\usepackage{bm,color,graphicx,amsmath,txfonts}
\usepackage{here}
\usepackage{array}
\usepackage{graphicx}
\usepackage[colorlinks=true,
citecolor=blue,
linkcolor=blue,
urlcolor=blue]{hyperref}
\usepackage{braket}
\usepackage{dsfont}
\usepackage[left=	2cm,top=2cm,right=1.2cm,bottom=2cm]{geometry}
\usepackage{tikz}
\usetikzlibrary{decorations.pathmorphing, positioning}

\everymath{\displaystyle}

\makeatletter
\renewcommand{\fnum@figure}{\textbf{Fig.~\thefigure:}}
\makeatother

\makeatletter
\renewcommand{\@makecaption}[2]{%
	\textbf{#1} #2\par
}
\makeatother
\makeatletter
\renewcommand{\fnum@figure}{\normalsize\textbf{Fig.~\thefigure:}}
\makeatother
\makeatletter
\renewcommand{\@makecaption}[2]{%
	\begin{flushleft}
		\textbf{#1} #2
	\end{flushleft}
}
\makeatother

\begin{document}
\makeatletter
\renewcommand{\@biblabel}[1]{%
	\makebox[2.1em][l]{\fontsize{10}{13}\selectfont[#1]}}
\makeatother

\title{Quantum Metrology and Quantum Battery in a Hyperfine-Coupled Spin System
}

\author{Omar Bachain}
\address{LPHE-Modeling and Simulation, Faculty of Sciences, Mohammed V University in Rabat, Rabat, Morocco}

\author{Mohamed \surname{Amazioug} }
\email{m.amazioug@uiz.ac.ma}
\address{LPTHE-Department of Physics, Faculty of Sciences, Ibnou Zohr University, Agadir 80000, Morocco}

\author{Rachid Ahl Laamara}
\address{LPHE-Modeling and Simulation, Faculty of Sciences, Mohammed V University in Rabat, Rabat, Morocco}
\address{Centre of Physics and Mathematics, CPM, Faculty of Sciences, Mohammed V University in Rabat, Rabat, Morocco}
\date{\today}

\begin{abstract}
	We investigate multiparameter quantum metrology and the energetic properties of an effective electron--proton spin system inspired by the hyperfine structure of the hydrogen atom. Considering the external magnetic field $\mathcal B$ and the hyperfine coupling strength $\Lambda$ as unknown parameters, we derive the quantum Fisher information matrix and establish the corresponding quantum Cram\'er--Rao bounds for their simultaneous estimation. We analyze the estimation precision and the correlations between the two parameters, showing that the weak-commutativity condition is satisfied despite the noncommutativity of the associated symmetric logarithmic derivatives. Independently, we explore the energetic properties of the system under a unitary charging protocol, characterizing its ergotropy, anti-ergotropy, storage capacity, and energy-basis coherence. We investigate the dependence of these quantities on the magnetic field, hyperfine coupling, temperature, and charging time. Our results provide a comprehensive characterization of the metrological properties and energetic performance of a hyperfine spin-based quantum system, highlighting its potential as a platform for quantum sensing and energy storage.
\end{abstract}

\maketitle

\section{Introduction}

The development of quantum technologies has stimulated considerable interest in understanding how quantum systems can be employed to store, transfer, and manipulate energy at the microscopic scale. Within this context, quantum thermodynamics provides a natural framework for investigating the energetic properties of finite-dimensional quantum systems, where coherence, correlations, and quantum fluctuations may play a decisive role \cite{Goold2016,Lostaglio2017,Vinjanampathy2016,Campaioli2024}. In particular, quantum batteries have emerged as promising quantum devices capable of storing and releasing energy through controlled quantum processes \cite{Binder2015,Campaioli2017,Le2018,Campaioli2024,Ferraro2026}. Unlike conventional energy-storage devices, their operation is governed by quantum dynamics and can therefore exhibit effects associated with coherence, entanglement, and collective interactions. In this regard, several studies have demonstrated that collective quantum effects and quantum correlations can influence the charging power and extractable work of quantum batteries \cite{AlickiFannes2013,Binder2015,Campaioli2017,Shi2022,Hovhannisyan2013}. Experimental investigations using NMR spin systems have also demonstrated the feasibility of implementing quantum battery charging protocols in controlled spin-based platforms \cite{JoshiMahesh2022}. Understanding the role of these quantum resources, as well as identifying the physical conditions under which they can be exploited, remains an important problem in quantum thermodynamics.

A fundamental quantity in this context is the ergotropy, introduced by Allahverdyan, Balian, and Nieuwenhuizen \cite{Allahverdyan2004}, which measures the maximum amount of work extractable from a quantum state through cyclic unitary transformations. Its value depends on both the energy spectrum of the system Hamiltonian and the eigenvalue distribution of the density operator. The concept of passivity provides the thermodynamic basis for this characterization, distinguishing states from which no work can be extracted through unitary operations from those possessing extractable energy \cite{PuszWoronowicz1978,Lenard1978}. Subsequent studies have investigated the influence of quantum coherence, correlations, and collective interactions on work extraction and battery charging \cite{AlickiFannes2013,Campaioli2017,Francica2017,Cakmak2020,Francica2020,Shi2022}. In particular, coherence in the energy eigenbasis can constitute an additional resource contributing to the extractable work \cite{Cakmak2020,Francica2020}. More recently, the concept of quantum battery capacity has been introduced as a complementary figure of merit quantifying the energy range accessible through unitary transformations \cite{Yang2023}. Further studies have investigated the optimization of ergotropy and storage capacity in interacting spin-chain quantum batteries \cite{AlKuwari2024}. Defined as the difference between the maximum and minimum energies attainable under unitary operations, this capacity characterizes the intrinsic energy-storage potential determined by the spectrum of the battery state and the energy spectrum of its Hamiltonian. Unlike the instantaneous ergotropy, it remains invariant under unitary charging, since such dynamics preserves the eigenvalues of the density operator \cite{Yang2023}. Its connections with entropy, coherence, and entanglement have further
motivated investigations of the energetic resources available in
finite-dimensional quantum systems  \cite{Yang2023,Campaioli2024,Zheng2026}. These developments emphasize that a comprehensive characterization of a quantum battery requires more than evaluating its stored energy alone.

Among the possible physical realizations of finite-dimensional quantum systems, interacting spin-$1/2$ particles provide a particularly useful setting because their energy spectra and quantum states can be controlled through external fields and spin--spin interactions. In this regard, the hydrogen atom offers a natural and physically meaningful example of a bipartite spin system, whose relevant degrees of freedom are associated with the electron and proton spins. In its electronic ground state, their interaction is governed by the hyperfine coupling, while an external magnetic field modifies the corresponding Zeeman energy levels and the hyperfine-state structure \cite{BreitMeyer1947,Antihydrogen2017,Diermaier2017}. The interplay between these two contributions determines the structure of the energy eigenstates and the thermal properties of the effective spin system. Hyperfine interactions have long been recognized as an important mechanism governing electron-spin dynamics in coupled spin systems and related nanostructures \cite{Schliemann2003}. More recently, the hyperfine-structure states of atomic hydrogen have been shown to exhibit natural and magnetically induced entanglement, demonstrating that an external magnetic field can modify and sustain quantum correlations against thermal effects \cite{Maleki2021}. These results make the electron--proton spin system an interesting platform for exploring the connection between microscopic spin interactions, quantum correlations, and energy-related properties. In particular, the competition between the hyperfine interaction and the Zeeman coupling provides two physically meaningful parameters, namely the magnetic field $\mathcal B$ and the hyperfine interaction strength $\Lambda$, which can be used to control and characterize the system.

In parallel with these developments, quantum metrology has established itself as a fundamental framework for determining the ultimate precision limits of physical-parameter estimation. The quantum Fisher information (QFI), which is closely related to the statistical distinguishability and geometrical structure of quantum states, provides a quantitative measure of the information encoded in a quantum state and determines the corresponding quantum Cram\'er--Rao bound \cite{BraunsteinCaves1994,Paris2009}. Its matrix generalization, the quantum Fisher information matrix (QFIM), is particularly relevant when several physical parameters must be estimated simultaneously \cite{Liu2020,Szczykulska2016}. In such situations, the estimation problem becomes more involved because the information associated with different parameters may be correlated and their optimal measurements may be incompatible \cite{Previdi2026}. Consequently, the individual quantum Cram\'er--Rao bounds do not, in general, provide a complete description of the attainable precision in a multiparameter estimation problem, where trade-offs between different parameter estimates may arise \cite{Ragy2016,Holevo2011}. The compatibility properties of the symmetric logarithmic derivatives (SLDs), and in particular the weak-commutativity condition, therefore play an essential role in assessing the attainability of the multiparameter quantum Cram\'er--Rao bound \cite{Ragy2016,Matsumoto2002}. Furthermore, recent studies have revealed fundamental connections between quantum Fisher information and quantum thermodynamics, showing that the QFI associated with time translations admits an operational interpretation in terms of the energetic coherence resources required for quantum-state preparation \cite{Marvian2022}. These observations motivate the use of multiparameter quantum metrology as a tool for characterizing interacting quantum systems beyond conventional parameter estimation.

Despite the substantial progress in quantum batteries and multiparameter quantum metrology \cite{Campaioli2024,Liu2020}, an important question remains concerning the precision with which the energetic properties of a quantum battery can themselves be inferred. In particular, evaluating the ergotropy or storage capacity as a function of physical parameters does not, by itself, quantify the uncertainty associated with their estimation when these parameters are experimentally unknown. This issue becomes especially relevant when the energetic quantities depend simultaneously on several unknown parameters, as commonly encountered in multiparameter quantum estimation \cite{Paris2009,Szczykulska2016}. Their estimation uncertainties are generally coupled through the off-diagonal elements of the inverse quantum Fisher information matrix (QFIM), which characterize the interplay between the different estimated parameters \cite{Ragy2016,Liu2020}. Consequently, the propagation of parameter-estimation uncertainties to nonlinear physical quantities represents an essential step in assessing the precision of experimentally inferred energetic observables \cite{Paris2009}. In this context, the QFIM provides a natural framework for establishing a connection between multiparameter estimation precision and the uncertainty bounds of energetic quantities, through the quantum Cram\'er--Rao bound and the corresponding error-propagation relations. Establishing this connection complements the conventional analysis of quantum battery performance by addressing not only how much work can be extracted or stored, but also how accurately these energetic quantities can be inferred from measurements.

Motivated by these considerations, in this work we investigate the multiparameter quantum metrology and energetic properties of an electron--proton spin system inspired by the hyperfine structure of the hydrogen atom \cite{Maleki2021}. The system is described by an isotropic hyperfine interaction of strength $\Lambda$, together with an external magnetic field $\mathcal B$, which couples differently to the electron and proton magnetic moments. Starting from the corresponding thermal Gibbs state, we first establish the spectral and thermodynamic properties of the system and investigate the simultaneous estimation of $\mathcal B$ and $\Lambda$ through the quantum Fisher information matrix (QFIM) \cite{Liu2020}. Using the vectorization formalism, we obtain the associated symmetric logarithmic derivatives (SLDs) and determine the quantum Cram\'er--Rao bounds for both parameters \cite{Paris2009}. We further examine the compatibility of their simultaneous estimation through the weak-commutativity condition \cite{Ragy2016} and analyze the determinant of the QFIM and the normalized off-diagonal information coefficient, thereby characterizing the joint information content and the correlations between the two estimation parameters.

Independently, we investigate the same electron--proton system as a quantum battery initially prepared in thermal equilibrium and subjected to a controlled transverse charging field \cite{Campaioli2017,Campaioli2024}. Within the considered unitary charging protocol, we characterize its energetic performance through the ergotropy, anti-ergotropy, and storage capacity \cite{Allahverdyan2004,Yang2023}, together with the $\ell_1$-norm of coherence in the energy eigenbasis. Their dependence on the charging time, magnetic field, hyperfine coupling, and initial temperature is analyzed in detail. To further assess the role of energy-basis coherence, we compare the energetic properties of the original charged state with those of its energy-dephased counterpart. This comparison allows us to investigate how removing energy-basis coherence modifies the extractable work and storage capacity.

We then investigate the same electron--proton system as a quantum battery initially prepared in thermal equilibrium and subjected to a controlled transverse charging field \cite{Campaioli2017,Campaioli2024}. Within the considered unitary charging protocol, we characterize its energetic performance through the ergotropy, anti-ergotropy, and storage capacity \cite{Allahverdyan2004,Yang2023}, together with the $\ell_1$-norm of coherence in the energy eigenbasis. Their dependence on the charging time, magnetic field, hyperfine coupling, and initial temperature is analyzed in detail. To further assess the role of energy-basis coherence, we compare the energetic properties of the original charged state with those of its energy-dephased counterpart. This comparison allows us to investigate how removing energy-basis coherence modifies the extractable work and storage capacity, complementing previous studies on the role of coherence in quantum work extraction \cite{Francica2017,Francica2020,Cakmak2020}.

Finally, we establish a quantitative connection between the multiparameter estimation problem and the energetic characterization of the battery by propagating the QFIM-based estimation bounds for $\mathcal B$ and $\Lambda$ to the ergotropy and storage capacity. Within the first-order error-propagation framework, this procedure provides lower bounds on the uncertainties associated with these energetic quantities and incorporates the correlations between the two estimated parameters \cite{Paris2009,Liu2020,Ragy2016}. We investigate how the resulting uncertainty bounds depend on the magnetic field, hyperfine coupling, and temperature, as well as on the charging time when considering the ergotropy. In contrast, the uncertainty bound associated with the storage capacity remains independent of the charging time under the considered unitary protocol. In this way, our analysis goes beyond the conventional evaluation of quantum battery performance by addressing not only how much work can be extracted or stored, but also how accurately these quantities can be inferred. It thus provides a unified framework in which the energetic properties and the metrological limitations of the same interacting quantum system are examined on an equal footing.

The paper is organized as follows. Section \ref{sec:model}
introduces the physical model and its thermal properties.
Section \ref{sec:QFIM} addresses multiparameter quantum
metrology. Section \ref{sec:QB} investigates the energetic
properties of the quantum battery, while Sec. \ref{sec:cc}
concludes the paper.
\section{Theoretical Model and Thermal State}
\label{sec:model}

We consider a bipartite quantum system composed of an electron and a proton, as realized in a hydrogen atom. Both constituents are described as spin-$1/2$ particles, and their mutual interaction is governed by the hyperfine coupling. In the presence of an external magnetic field, the system exhibits a competition between the spin--spin interaction and the Zeeman coupling. This system provides a suitable framework for investigating the interplay between quantum thermodynamics, quantum correlations, and multiparameter quantum metrology.

\begin{figure}[t]
\includegraphics[width=1\linewidth]{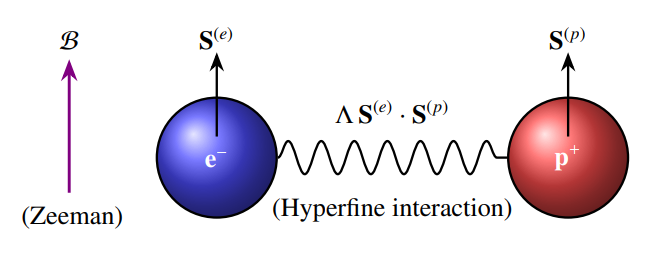}
	\caption{Schematic representation of the electron--proton
		spin system under an external magnetic field. The electron
		and proton spins are coupled through the hyperfine
		interaction.}
	\label{figure1}
\end{figure}

The Hamiltonian describing the electron--proton system is given by
\begin{equation}
	H=\Lambda\,\mathbf{S}^{(e)}\cdot\mathbf{S}^{(p)}
	+B\left(\mu S_z^{(e)}-\nu S_z^{(p)}\right),
	\label{eq:Hamiltonian}
\end{equation}
where $\mathbf{S}^{(e)}$ and $\mathbf{S}^{(p)}$ denote the spin operators of the electron and proton, respectively. The parameter $\Lambda$ characterizes the hyperfine interaction, while $B$ represents the magnitude of the applied magnetic field. The quantities $\mu$ and $\nu$ denote the magnetic coupling constants of the electron and proton, respectively. Their magnitudes satisfy approximately $\nu\simeq\mu/1836.14$, reflecting the large difference between the electron and proton masses. The opposite signs in the Zeeman contribution account for the opposite magnetic moments associated with the electron and proton spin degrees of freedom. The hyperfine interaction provides the physical basis for describing the coupling between the two spin degrees of freedom \cite{Erlingsson2002,Schliemann2003,Maleki2021}. The physical configuration of the system described by
Eq.~(\ref{eq:Hamiltonian}) is schematically illustrated
in Fig.~\ref{figure1}.

Throughout this work, we use natural units, $\hbar=k_{\mathrm B}=1$. Since both constituents are spin-$1/2$ particles, their spin operators can be expressed in terms of the Pauli matrices as
\begin{equation}
	\mathbf{S}^{(j)}=\frac{1}{2}\boldsymbol{\sigma}^{(j)},
	\qquad j=e,p,
	\label{eq:spinoperators}
\end{equation}
where $\boldsymbol{\sigma}^{(j)}=(\sigma_x^{(j)},\sigma_y^{(j)},\sigma_z^{(j)})$. We adopt the computational basis
\begin{equation}
	\mathcal{B}=
	\left\{
	|00\rangle,|01\rangle,|10\rangle,|11\rangle
	\right\},
	\label{eq:computationalbasis}
\end{equation}
with the convention $|0\rangle=|\uparrow\rangle$ and $|1\rangle=|\downarrow\rangle$.

In this basis, the Hamiltonian takes the matrix representation
\begin{equation}
	H=
	\begin{pmatrix}
		\frac{\Lambda}{4}+\delta&0&0&0\\
		0&-\frac{\Lambda}{4}+\Delta&\frac{\Lambda}{2}&0\\
		0&\frac{\Lambda}{2}&-\frac{\Lambda}{4}-\Delta&0\\
		0&0&0&\frac{\Lambda}{4}-\delta
	\end{pmatrix},
	\label{eq:Hmatrix}
\end{equation}
where we have introduced the compact notations
\begin{equation}
	\delta=\frac{B}{2}(\mu-\nu),
	\qquad
	\Delta=\frac{B}{2}(\mu+\nu).
	\label{eq:deltaDelta}
\end{equation}
The matrix structure in Eq.~(\ref{eq:Hmatrix}) shows that the states $|00\rangle$ and $|11\rangle$ are eigenstates of the Hamiltonian, whereas the two-dimensional subspace spanned by $\{|01\rangle,|10\rangle\}$ is coupled by the hyperfine interaction. Consequently, the diagonalization of the full Hamiltonian reduces to the diagonalization of this coupled subspace. Introducing
\begin{equation}
	\Omega=\sqrt{\Delta^2+\frac{\Lambda^2}{4}},
	\label{eq:Omega}
\end{equation}
the four eigenvalues are obtained as
\begin{equation}
	E_1=\frac{\Lambda}{4}+\delta,
	\qquad
	E_2=\frac{\Lambda}{4}-\delta,
	\label{eq:E12}
\end{equation}
and
\begin{equation}
	E_3=-\frac{\Lambda}{4}-\Omega,
	\qquad
	E_4=-\frac{\Lambda}{4}+\Omega.
	\label{eq:E34}
\end{equation}

The corresponding normalized eigenvectors are
\begin{equation}
	|E_1\rangle=|00\rangle,
	\qquad
	|E_2\rangle=|11\rangle,
	\label{eq:eigenvectors12}
\end{equation}
and
\begin{equation}
	|E_3\rangle=
	\frac{\alpha_-|01\rangle+|10\rangle}
	{\sqrt{1+\alpha_-^2}},
	\qquad
	|E_4\rangle=
	\frac{\alpha_+|01\rangle+|10\rangle}
	{\sqrt{1+\alpha_+^2}},
	\label{eq:eigenvectors34}
\end{equation}
where
\begin{equation}
	\alpha_\pm=\frac{2(\Delta\pm\Omega)}{\Lambda}.
	\label{eq:alphapm}
\end{equation}
These expressions are valid for $\Lambda\neq0$. The eigenvectors associated with $E_3$ and $E_4$ describe coherent superpositions of the electron--proton spin configurations $|01\rangle$ and $|10\rangle$. Their mixing is controlled by the competition between the hyperfine coupling and the external magnetic field. In particular, the relation
\begin{equation}
	\alpha_-\alpha_+=-1,
	\label{eq:alpharel}
\end{equation}
ensures the orthogonality of these two eigenstates.

We assume that the electron--proton system is initially prepared in thermal equilibrium at temperature $T$. The corresponding Gibbs state is defined as
\begin{equation}
	\rho_{\mathrm{th}}=
	\frac{e^{-H/T}}{Z},
	\label{eq:Gibbs}
\end{equation}
where
\begin{equation}
	Z=\operatorname{Tr}\left(e^{-H/T}\right)
	\label{eq:partitiondefinition}
\end{equation}
is the partition function. Using the spectral decomposition of the Hamiltonian, the thermal state can equivalently be expressed as
\begin{equation}
	\rho_{\mathrm{th}}=
	\frac{1}{Z}\sum_{n=1}^{4}
	e^{-E_n/T}|E_n\rangle\langle E_n|.
	\label{eq:spectralGibbs}
\end{equation}

Substituting the eigenvalues given in Eqs.~(\ref{eq:E12}) and (\ref{eq:E34}) into Eq.~(\ref{eq:partitiondefinition}), we obtain the partition function
\begin{equation}
	Z=
	2e^{-\Lambda/(4T)}
	\cosh\left(\frac{\delta}{T}\right)
	+
	2e^{\Lambda/(4T)}
	\cosh\left(\frac{\Omega}{T}\right).
	\label{eq:partitionfunction}
\end{equation}

In the computational basis, the thermal density matrix assumes an $X$-state structure,
\begin{equation}
	\rho_{\mathrm{th}}=
	\begin{pmatrix}
		\rho_{11}&0&0&0\\
		0&\rho_{22}&\rho_{23}&0\\
		0&\rho_{23}&\rho_{33}&0\\
		0&0&0&\rho_{44}
	\end{pmatrix},
	\label{eq:thermalmatrix}
\end{equation}
where the nonzero matrix elements are given by
\begin{align}
	\rho_{11}&=	\frac{1}{Z}
	\exp\left[-\frac{1}{T}
	\left(\frac{\Lambda}{4}+\delta\right)\right],
	\label{eq:rho11}
\\
	\rho_{44}&=	\frac{1}{Z}
	\exp\left[-\frac{1}{T}
	\left(\frac{\Lambda}{4}-\delta\right)\right],
	\label{eq:rho44}
\\
	\rho_{22}&=
	\frac{e^{\Lambda/(4T)}}{Z}
	\left[
	\cosh\left(\frac{\Omega}{T}\right)
	-\frac{\Delta}{\Omega}
	\sinh\left(\frac{\Omega}{T}\right)
	\right],
	\label{eq:rho22}
\\
	\rho_{33}&=
\frac{	e^{\Lambda/(4T)}}{Z}
	\left[
	\cosh\left(\frac{\Omega}{T}\right)
	+\frac{\Delta}{\Omega}
	\sinh\left(\frac{\Omega}{T}\right)
	\right],
	\label{eq:rho33}
\\
	\rho_{23}&=
	-\frac{\Lambda}{2Z\Omega}
	e^{\Lambda/(4T)}
	\sinh\left(\frac{\Omega}{T}\right).
	\label{eq:rho23}
\end{align}
The remaining off-diagonal elements vanish. The matrix elements given in Eqs.~(\ref{eq:rho11})--(\ref{eq:rho23}) already include the normalization factor $1/Z$, ensuring that the thermal density operator satisfies $\operatorname{Tr}(\rho_{\mathrm{th}})=1$.

The structure of Eq.~(\ref{eq:thermalmatrix}) is particularly relevant to the subsequent analysis. The diagonal entries describe the thermal populations of the computational-basis states, whereas the off-diagonal element $\rho_{23}$ characterizes the coherence between the states $|01\rangle$ and $|10\rangle$. This coherence originates from the hyperfine interaction and reflects the mixing of the two spin configurations in the energy eigenbasis. Its magnitude depends on both the interaction strength and the temperature, thereby providing a direct link between the microscopic coupling and the quantum properties of the thermal state.

At finite temperature, the density operator generally represents a mixed state resulting from the thermal occupation of several energy levels. In the low-temperature limit, the thermal population becomes increasingly concentrated in the ground state, provided that the ground state is nondegenerate. Conversely, in the high-temperature limit, the Gibbs state approaches the maximally mixed state,
\begin{equation}
	\lim_{T\rightarrow\infty}\rho_{\mathrm{th}}
	=\frac{\mathbb{I}_4}{4},
	\label{eq:highTlimit}
\end{equation}
where $\mathbb{I}_4$ denotes the identity operator in the two-qubit Hilbert space. These limiting behaviors provide useful consistency checks for the thermal-state expressions and establish temperature as a central parameter governing the equilibrium properties of the system \cite{MillerAnders2018}.

The thermal density matrix derived above constitutes the initial state used in the subsequent investigation of the quantum battery characteristics and multiparameter estimation performance of the electron--proton system.

\section{Quantum Fisher Information Matrix}
\label{sec:QFIM}

We investigate the multiparameter quantum estimation of the external
magnetic field $\mathcal B$ and the hyperfine interaction strength
$\Lambda$ in the electron--proton system. The ultimate precision
limits associated with their simultaneous estimation are quantified
by the QFIM. To evaluate this
matrix, we employ the vectorization formalism, which provides an
efficient computational implementation of the standard symmetric
logarithmic derivative (SLD) approach without requiring the explicit
diagonalization of the density operator
\cite{vsafranek2018simple,Matsumoto2002,crowley2014tradeoff}.
It should be emphasized that this formalism does not introduce a
new metrological framework, but rather offers a convenient
representation of the conventional SLD equations.

Let $\rho(\mathcal B,\Lambda)$ denote the density operator of the
system, which depends on the two unknown parameters
$\boldsymbol{\theta}=(\mathcal B,\Lambda)$. The QFIM is defined as
\begin{equation}
	\mathcal N_{\mu\nu}
	=
	\frac{1}{2}\operatorname{Tr}
	\left[
	\rho\left(
	\mathsf L_\mu\mathsf L_\nu+
	\mathsf L_\nu\mathsf L_\mu
	\right)
	\right],
	\label{eq:QFIM_definition}
\end{equation}
where $\mathsf L_\mu$ is the SLD associated with the parameter
$\theta_\mu$. These operators are determined by \cite{Carollo2018}
\begin{equation}
	\frac{\partial\rho}{\partial\theta_\mu}
	=
	\frac{1}{2}
	\left(
	\mathsf L_\mu\rho+\rho\mathsf L_\mu
	\right).
	\label{eq:SLD_definition}
\end{equation}
For $M$ independent and identically prepared copies of the
quantum state, the quantum Cram\'er--Rao inequality reads \cite{Paris2009}
\begin{equation}
	\operatorname{Cov}(\hat{\boldsymbol{\theta}})
	\succeq \frac{1}{M}\mathcal N^{-1},
	\label{eq:QCRB}
\end{equation}
In the present analysis, we set $M=1$ and consider the QFIM associated with a single copy of the quantum state. Accordingly, the inverse QFIM provides the single-copy quantum Cram\'er--Rao lower bound on the covariance matrix of the parameter estimators.

For an arbitrary operator $A$ acting on an $n$-dimensional Hilbert
space, its vectorized representation is defined as \cite{gilchrist2009vectorization,Helstrom1976}
\begin{equation}
	|A\rangle\rangle
	=
	\sum_{k,l}a_{kl}|l\rangle\otimes|k\rangle,
	\label{eq:vectorization}
\end{equation}
where $a_{kl}=\langle k|A|l\rangle$. This mapping corresponds to
stacking the columns of the matrix representation of $A$ into a
single column vector. It satisfies the identities
\begin{align}
	(B\otimes A)|C\rangle\rangle
	&=|ACB^T\rangle\rangle,
	\label{eq:vec_identity1}\\
	\langle\langle A|B\rangle\rangle
	&=\operatorname{Tr}(A^\dagger B).
	\label{eq:vec_identity2}
\end{align}
Applying this formalism to Eq.~(\ref{eq:SLD_definition}), we obtain
\begin{equation}
	\left(\rho^T\otimes I+I\otimes\rho\right)
	|\mathsf L_\mu\rangle\rangle
	=
	2\left|
	\frac{\partial\rho}{\partial\theta_\mu}
	\right\rangle\rangle.
	\label{eq:vectorized_SLD}
\end{equation}
Introducing the superoperator
\begin{equation}
	\mathcal P=\rho^T\otimes I+I\otimes\rho,
	\label{eq:superoperator}
\end{equation}
the vectorized SLD operators can be expressed as
\begin{equation}
	\big|\mathsf L_\mu\big\rangle\big\rangle
	=
	2\mathcal P^{-1}
	\Bigg|
	\frac{\partial\rho}{\partial\theta_\mu}
	\Bigg\rangle\Bigg\rangle,
	\label{eq:SLD_vectorized}
\end{equation}
where $\mathcal P^{-1}$ denotes the inverse of the superoperator
$\mathcal P$, which is assumed to be nonsingular.
Accordingly, the QFIM can be rewritten in the compact form
\begin{equation}
\mathcal N_{\mu\nu}
	=
	2\left\langle\!\left\langle
	\frac{\partial\rho}{\partial\theta_\mu}
	\right|
	\mathcal P^{-1}
	\left|
	\frac{\partial\rho}{\partial\theta_\nu}
	\right\rangle\!\right\rangle.
	\label{eq:QFIM_vectorized}
\end{equation}
This expression allows the QFIM to be evaluated directly from the
density operator and its parameter derivatives.

For the two-qubit electron--proton system, expressed in the
computational basis
$\{|00\rangle,|01\rangle,|10\rangle,|11\rangle\}$, the vectorized
derivatives with respect to $\mathcal B$ and $\Lambda$ are given by
\begin{align}
	|\partial_{\mathcal B}\rho\rangle\rangle
	={}&
	\Big(
	\partial_{\mathcal B}\rho_{11},0,0,0,\,
	0,\partial_{\mathcal B}\rho_{22},
	\partial_{\mathcal B}\rho_{23},0,\nonumber\\
	&\quad
	0,\partial_{\mathcal B}\rho_{23},
	\partial_{\mathcal B}\rho_{33},0,\,
	0,0,0,\partial_{\mathcal B}\rho_{44}
	\Big)^T,
	\label{eq:derivative_B}\\
	|\partial_{\Lambda}\rho\rangle\rangle
	={}&
	\Big(
	\partial_{\Lambda}\rho_{11},0,0,0,\,
	0,\partial_{\Lambda}\rho_{22},
	\partial_{\Lambda}\rho_{23},0,\nonumber\\
	&\quad
	0,\partial_{\Lambda}\rho_{23},
	\partial_{\Lambda}\rho_{33},0,\,
	0,0,0,\partial_{\Lambda}\rho_{44}
	\Big)^T.
	\label{eq:derivative_Lambda}
\end{align}
Consequently, by applying Eq.~(\ref{eq:SLD_vectorized}) to the parameter derivatives given in Eqs.~(\ref{eq:derivative_B}) and (\ref{eq:derivative_Lambda}), we obtain the symmetric logarithmic derivative (SLD) operators associated with the estimation of the external magnetic field $\mathcal B$ and the hyperfine interaction strength $\Lambda$. For convenience, we first introduce the following auxiliary quantities
\begin{align}
	C&=\cosh\left(\frac{\Omega}{T}\right),
	\quad
	S=\sinh\left(\frac{\Omega}{T}\right),
	\quad J=\tanh\left(\frac{\Omega}{T}\right)\\
	&\quad D=
	e^{\frac{3\delta+\Delta}{2T}}
	+e^{\frac{\Delta-\delta}{2T}}
	+2e^{\frac{\Lambda+\delta+\Delta}{2T}}C.
\end{align}
In the computational basis, the SLD operator
associated with the estimation of $\mathcal B$
takes the following $4\times4$ matrix form
\begin{equation}
	\mathsf L_{\mathcal B}
	=
	\begin{pmatrix}
		L_{11}^{(\mathcal B)}&0&0&0\\
		0&L_{22}^{(\mathcal B)}&	 L_{23}^{(\mathcal{B})}&0\\
		0&	 L_{23}^{(\mathcal{B})}&L_{33}^{(\mathcal B)}&0\\
		0&0&0&L_{44}^{(\mathcal B)}
	\end{pmatrix}.
	\label{eq:SLD_B}
\end{equation}
\begin{align}
	L_{11}^{(\mathcal B)}
	={}&
-\frac{2e^{\frac{\delta}{T}}}{BT\Omega D}
\left[
\delta\Omega e^{\frac{\delta}{T}}
+e^{\frac{\Lambda}{2T}}
\left(\delta\Omega C+\Delta^2S\right)
\right],
	\\\nonumber
	L_{22}^{(\mathcal B)}
	={}&
-\frac{1}{4BT\Omega^3D}
\Bigg\{
8\Delta^3\Omega
e^{\frac{\Lambda+2\delta}{2T}}C+T\Lambda^2\Delta
\left(1+e^{\frac{2\delta}{T}}\right)J
\\\nonumber
&+\Omega\Big[
e^{\frac{2\delta}{T}}
\left(\Lambda^2\delta+
4\Delta^2(\delta+\Delta)\right)
-\Lambda^2\delta
+4\Delta^2(\Delta-\delta)\\
&+2\Delta e^{\frac{\Lambda+2\delta}{2T}}
\left(T\Lambda^2+\Delta\Lambda^2+4\Delta^3\right)S
\Big]
\Bigg\},
	\\
	L_{33}^{(\mathcal B)}
	={}&\nonumber
	\frac{1}{4BT\Omega^{3}D}
	\Bigg\{
	8\Delta^{3}\Omega
	e^{\frac{\Lambda+2\delta}{2T}}C
+2\Delta e^{\frac{\Lambda+2\delta}{2T}}
	\left(T\Lambda^{2}-4\Delta\Omega^{2}\right)S
	\\\nonumber
	&+\Omega\Big[
	\Lambda^{2}\delta
	+4\Delta^{2}(\delta+\Delta)
	+e^{\frac{2\delta}{T}}
	\left(-\Lambda^{2}\delta
	+4\Delta^{2}(\Delta-\delta)\right)
	\Big]
	\\
	&+T\Lambda^{2}\Delta
	\left(1+e^{\frac{2\delta}{T}}\right)
	J
	\Bigg\},
	\\
		L_{23}^{(\mathcal B)}
	={}&
	-\frac{\Lambda\Delta^{2}}{2BT\Omega^{3}}
	\left[
	\Omega-TJ
	\right],
	\\
	L_{44}^{(\mathcal B)}
	={}&
	\frac{2}{BT\Omega D}
	\left[
	\delta\Omega
	+
	e^{\frac{\Lambda+2\delta}{2T}}
	\left(
	\delta\Omega C-\Delta^{2}S
	\right)
	\right].
\end{align}
Similarly, the SLD operator associated with the estimation of the hyperfine interaction strength $\Lambda$ takes the following matrix form

\begin{equation}
	\mathsf L_{\Lambda}
	=
	\begin{pmatrix}
		L_{11}^{(\Lambda)} & 0
		& 0 &0\\
		0 & L_{22}^{(\Lambda)}
		& L_{23}^{(\Lambda)} & 0\\
		0 & L_{23}^{(\Lambda)}
		& L_{33}^{(\Lambda)} & 0\\
	0& 0
		& 0 & L_{44}^{(\Lambda)}
	\end{pmatrix},
	\label{eq:SLD_Lambda}
\end{equation}
\begin{align}
			L_{11}^{(\Lambda)}
			={}&-\frac{
				e^{\frac{\Lambda+2\delta}{2T}}
				\left(2\Omega C+\Lambda S\right)
			}{
				2T\Omega D
			},
		\\\nonumber
			L_{22}^{(\Lambda)}
		={}&\frac{1}{T\Omega^{3}D}
		\Bigg\{
		-\frac{\Lambda\Delta\Omega}{2}
		e^{\frac{\Lambda+2\delta}{2T}}C
		-\frac{\Lambda}{2}
		\left(\Omega^{2}-T\Delta\right)
		e^{\frac{\Lambda+2\delta}{2T}}S
		\\
		&+\left(1+e^{\frac{2\delta}{T}}\right)
		\left[
		\frac{\Omega}{8}
		\left(\Lambda^{2}-2\Lambda\Delta+4\Delta^{2}\right)
		+\frac{T\Lambda\Delta}{4}
	J
		\right]
		\Bigg\},
		\\\nonumber
			L_{33}^{(\Lambda)}
		={}&\frac{1}{8T\Omega^{3}D}
		\Bigg\{
		4\Lambda\Delta\Omega
		e^{\frac{\Lambda+2\delta}{2T}}C
	-4\Lambda\left(\Omega^{2}+T\Delta\right)
		e^{\frac{\Lambda+2\delta}{2T}}S
		\\
		&+\left(1+e^{\frac{2\delta}{T}}\right)
		\Bigg[
		\Omega\left(\Lambda^{2}+2\Lambda\Delta+4\Delta^{2}\right)
		-2T\Lambda\Delta
			J
		\Bigg]
		\Bigg\},
		\\
			L_{23}^{(\Lambda)}
		={}&	-\frac{
			\Lambda^{2}\Omega
			+4T\Delta^{2}J
		}{
			8T\Omega^{3}
		},
		\\
		L_{44}^{(\Lambda)}
	={}&	-\frac{
			e^{\frac{\Lambda+2\delta}{2T}}
			\left(2\Omega C+\Lambda S\right)
		}{
			2T\Omega D
		}.
\end{align}
Substituting these operators into Eq.~(\ref{eq:QFIM_definition}),
the QFIM assumes the symmetric matrix form
\begin{equation}
\mathcal N=
	\begin{pmatrix}
	\mathcal N_{\mathcal B\mathcal B}
		&
	\mathcal N_{\mathcal B\Lambda}\\
	\mathcal N_{\mathcal B\Lambda}
		&
	\mathcal N_{\Lambda\Lambda}
	\end{pmatrix},
	\label{eq:QFIM_matrix}
\end{equation}
where its elements are given by
\begin{align}
\mathcal N_{\mathcal B\mathcal B}
	&=
	\operatorname{Tr}
	\left(\rho\mathsf L_{\mathcal B}^{\,2}\right),
	\label{eq:FBB}\\
\mathcal N_{\mathcal B\Lambda}
	&=
	\frac{1}{2}\operatorname{Tr}
	\left[
	\rho\left(
	\mathsf L_{\mathcal B}\mathsf L_\Lambda+
	\mathsf L_\Lambda\mathsf L_{\mathcal B}
	\right)
	\right],
	\label{eq:FBLambda}\\
\mathcal N_{\Lambda\Lambda}
	&=
	\operatorname{Tr}
	\left(\rho\mathsf L_\Lambda^{\,2}\right).
	\label{eq:FLambdaLambda}
\end{align}
The explicit analytical expressions of the three independent elements of the QFIM are provided in Appendix~\ref{app:QFIM}.

To further characterize the coupling between the two estimated parameters, we introduce the normalized off-diagonal QFIM coefficient,
\begin{equation}
\mathcal 	R_{\mathcal B\Lambda}
	=
	\frac{\mathcal N_{\mathcal B\Lambda}}
	{\sqrt{\mathcal N_{\mathcal B\mathcal B}
			\mathcal N_{\Lambda\Lambda}}}.
	\label{eq:metrological_correlation}
\end{equation}
This dimensionless quantity measures the relative contribution of the cross-information term to the QFIM. Its magnitude is bounded by unity, $|\mathcal R_{\mathcal B\Lambda}|\leq 1$, owing to the positive semidefiniteness of the QFIM. In particular, it provides insight into the degree of coupling between the two parameter-estimation processes. A vanishing value indicates the absence of a cross-information contribution, whereas a nonzero value reflects the presence of statistical coupling between the estimated parameters.

The determinant of the QFIM is
\begin{equation}
	\det\mathcal N
	=
\mathcal N_{\mathcal B\mathcal B}
\mathcal N_{\Lambda\Lambda}
	-
\mathcal N_{\mathcal B\Lambda}^{\,2}.
	\label{eq:det_QFIM}
\end{equation}
The resulting quantum Cram\'er--Rao bounds for the simultaneous
estimation of the magnetic field and the interaction strength read
\begin{align}
	(\Delta\mathcal B)^2
	&\geq
	\frac{\mathcal N_{\Lambda\Lambda}}{\det\mathcal N},
	\label{eq:bound_B}\\
	(\Delta\Lambda)^2
	&\geq
	\frac{\mathcal N_{\mathcal B\mathcal B}}{\det\mathcal N}.
	\label{eq:bound_Lambda}
\end{align}

To investigate the compatibility of the simultaneous estimation of the external magnetic field $\mathcal B$ and the hyperfine interaction strength $\Lambda$, we examine the commutativity properties of their corresponding symmetric logarithmic derivatives (SLDs) \cite{Albarelli2026}. In general, the commutativity of the SLD operators is a sufficient condition for the simultaneous attainability of the associated quantum Cram\'er--Rao bounds. However, strict commutativity is not necessary in the asymptotic regime, where the weaker condition of weak commutativity can also ensure the attainability of the SLD bound under collective measurements \cite{ragy2016compatibility,Chen2022}. 

Therefore, the weak-commutativity condition, $\operatorname{Tr}(\rho_{\mathrm{th}}[L_{\mathcal B},L_{\Lambda}])=0$, is satisfied, although the two SLD operators do not commute in general. This condition establishes the asymptotic compatibility of the multiparameter estimation problem under collective measurements on many identical copies of the quantum statev\cite{ragy2016compatibility,Chen2022}. It does not, however, imply the simultaneous attainability of the individual SLD quantum Cram\'er--Rao bounds with a single copy. Accordingly, the bounds considered in this work are interpreted as single-copy precision benchmarks, while their simultaneous asymptotic attainability follows from the weak-commutativity condition \cite{Zhou2026}.
	\begin{figure}[H]
	\centering
	\includegraphics[width=0.95\linewidth]{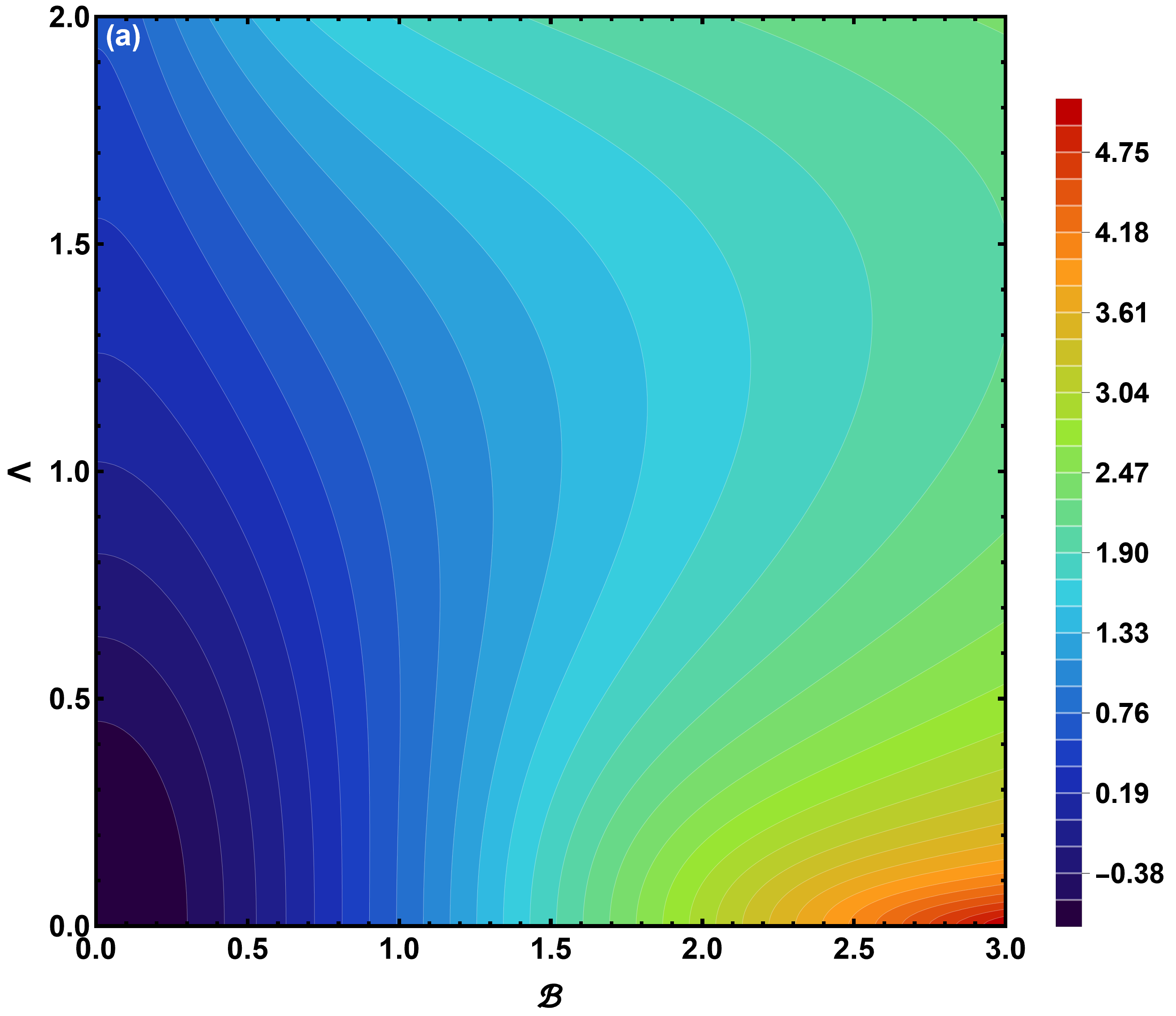}
	\includegraphics[width=0.95\linewidth]{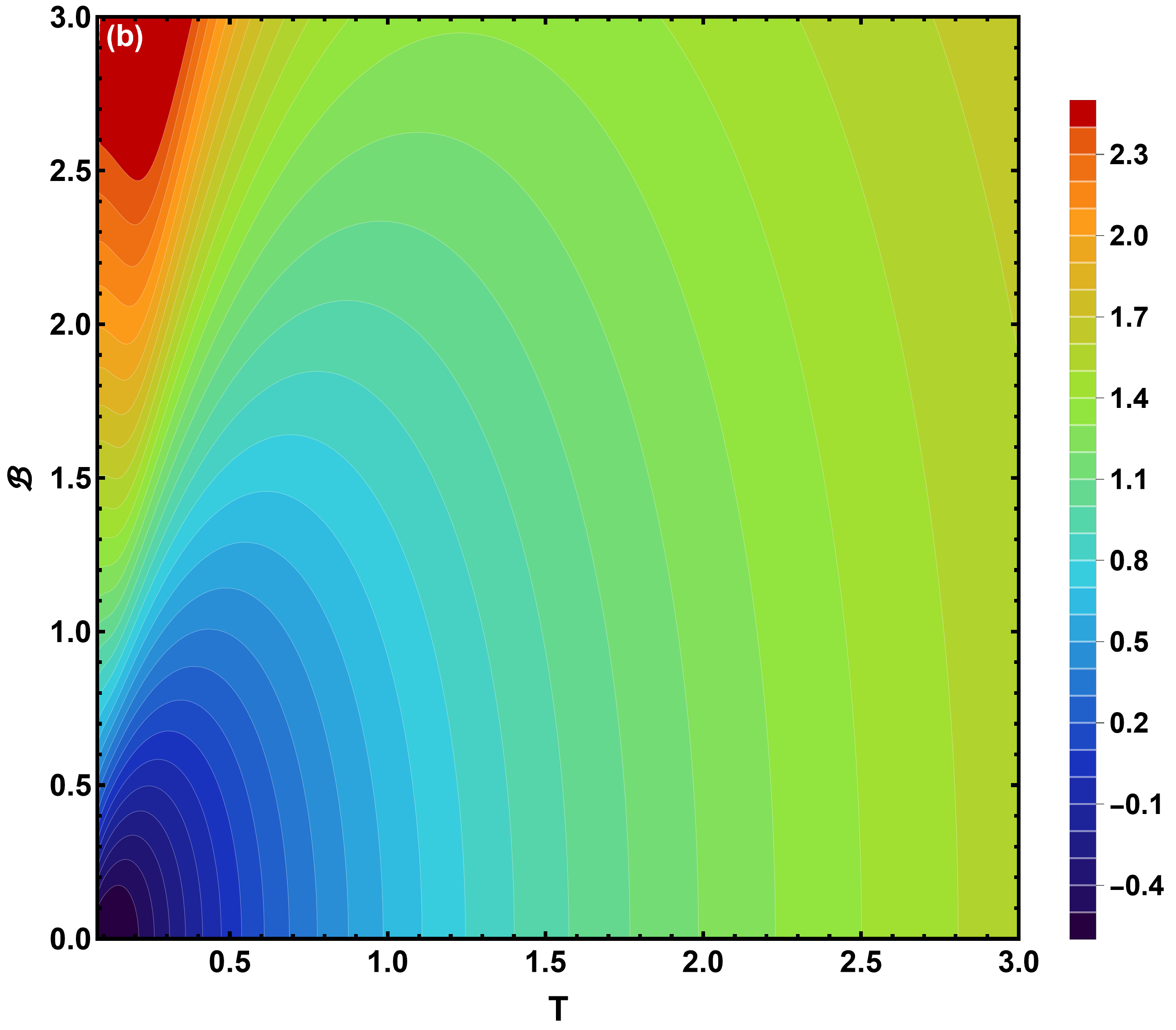}
\caption{Contour plots of the logarithm of the minimum variance, $\log_{10}[(\Delta\mathcal B)^2_{\min}]$, as a function of (a) $\mathcal B$ and $\Lambda$ at $T=0.2$, and (b) $\mathcal B$ and $T$ at $\Lambda=0.5$. The remaining parameters are $\mu=1$ and $\nu=\mu/1836.14$.}
	\label{figure2}
\end{figure}
	\begin{figure}
	\centering
	\includegraphics[width=0.95\linewidth]{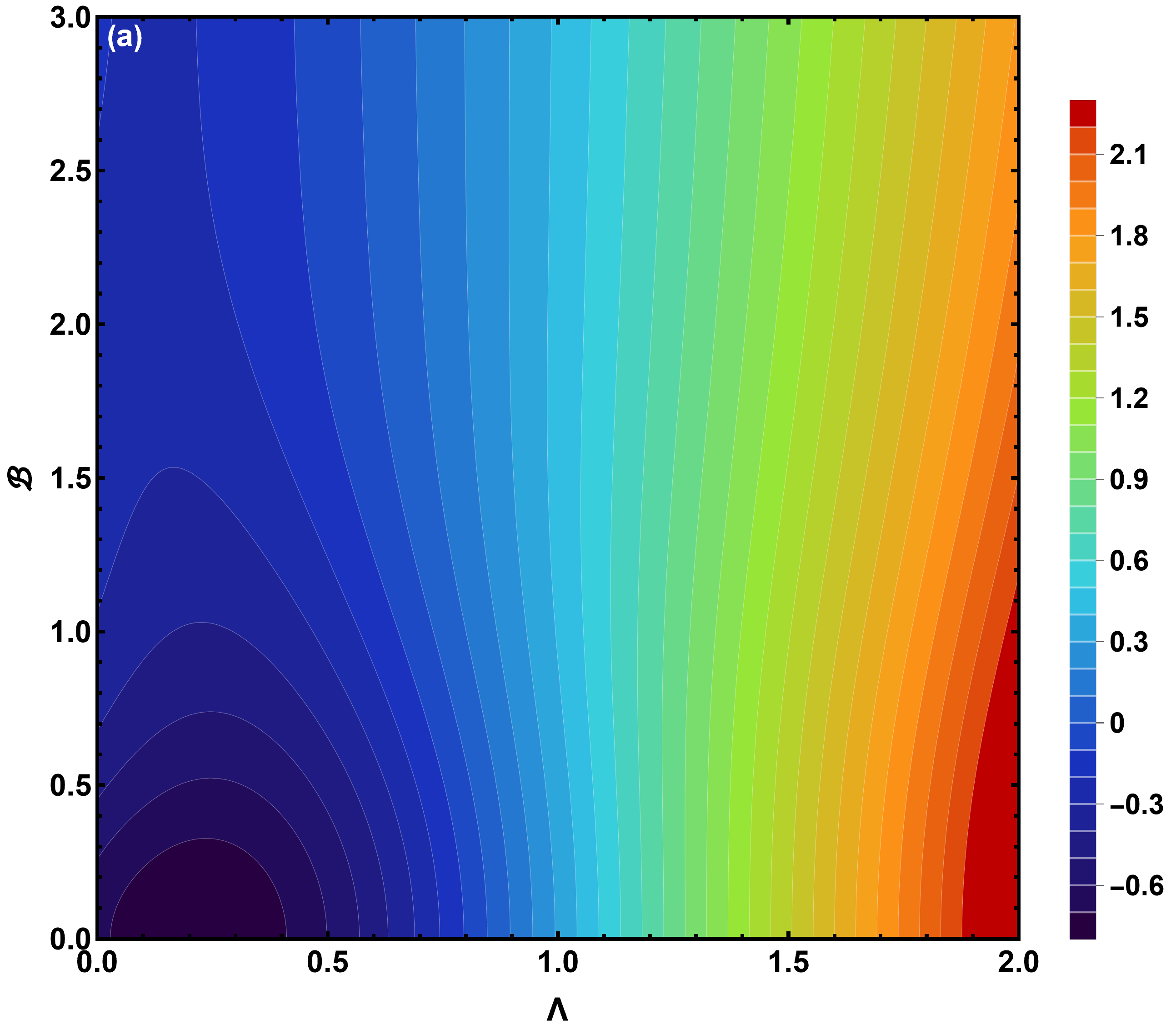}
	\includegraphics[width=0.95\linewidth]{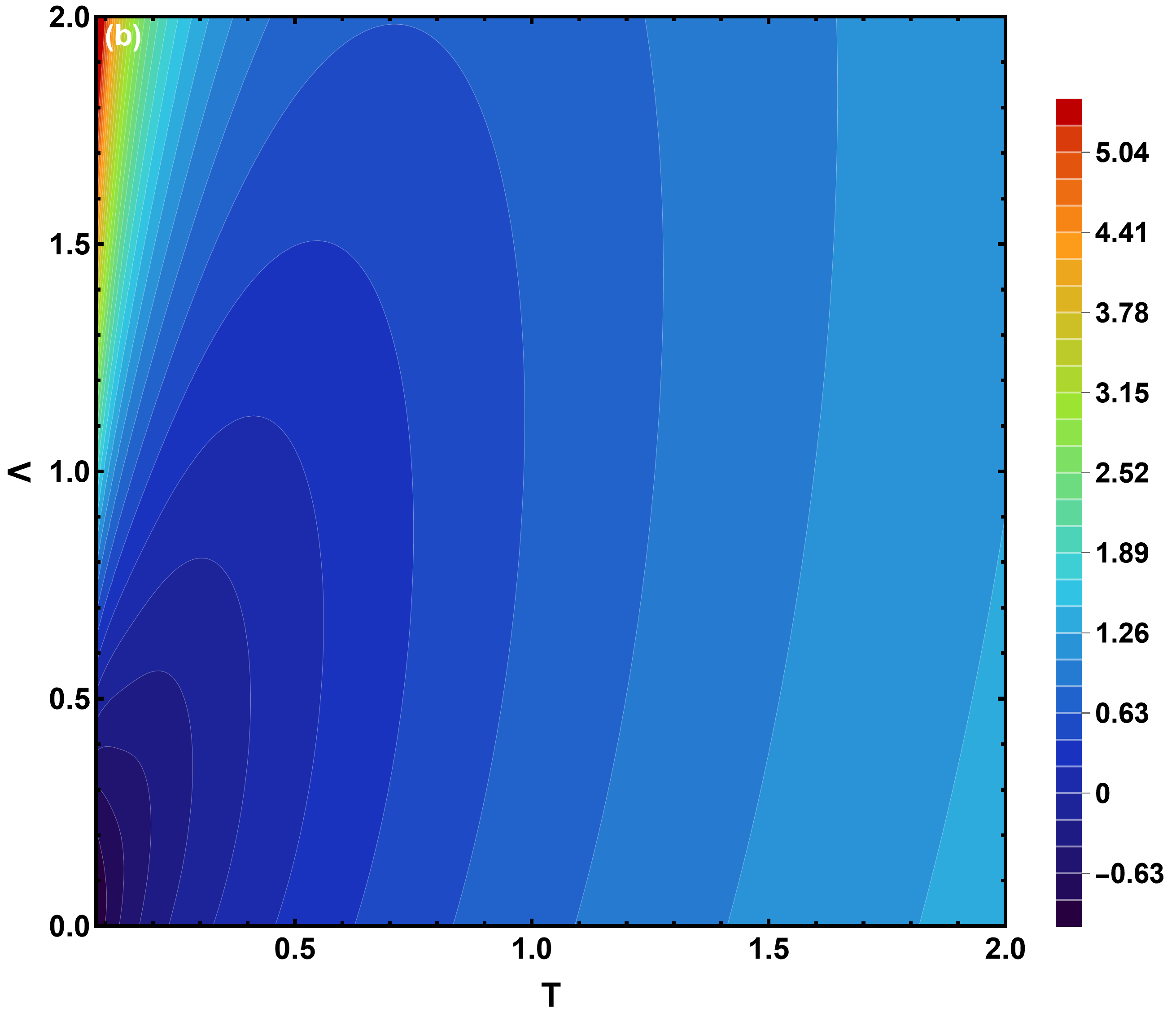}
	\caption{Contour plots of the logarithm of the minimum variance, $\log_{10}[(\Delta\Lambda)^2_{\min}]$, as a function of (a) $\Lambda$ and $\mathcal B$ at $T=0.2$, and (b) $\Lambda$ and $T$ at $\mathcal B=1$. The remaining parameters are $\mu=1$ and $\nu=\mu/1836.14$.}
	\label{figure3}
\end{figure}

Having established the multiparameter quantum Fisher information matrix and the corresponding quantum Cram\'er–Rao bounds, we now investigate how the attainable estimation precision depends on the physical parameters of the system. We begin by examining the minimum variance associated with the estimation of the external magnetic field \(\mathcal B\), considering the effects of the hyperfine coupling strength \(\Lambda\) and the temperature \(T\).

Figure \ref{figure2} illustrates the dependence of the minimum variance associated with the estimation of the magnetic field \(\mathcal B\) on the hyperfine coupling strength \(\Lambda\) and the temperature \(T\). As shown in Fig. \ref{figure2}(a), the estimation uncertainty generally increases with \(\mathcal B\), while its dependence on \(\Lambda\) reveals a nontrivial interplay between the two parameters. As follows from Eq. (\ref{eq:bound_B}), the attainable precision is governed not only by the information associated with \(\mathcal B\), but also by the information on \(\Lambda\) and the correlations between their estimations. In particular, the off-diagonal QFIM element \(\mathcal N_{\mathcal B\Lambda}\) accounts for the statistical coupling between the two parameters.

Figure \ref{figure2}(b) further highlights the role of temperature in magnetic-field estimation. The uncertainty exhibits a pronounced dependence on both \(\mathcal B\) and \(T\), with particularly large values in the low-temperature, high-field region. This behavior is closely related to the temperature-dependent thermal populations and the field-dependent energy spectrum. Indeed, the hyperfine interaction modifies the energy splitting $\Omega$, thereby affecting the parameter derivatives of the Gibbs state and, consequently, the QFIM. Overall, these results demonstrate that the attainable precision in estimating \(\mathcal B\) is governed by the combined effects of thermal mixing and hyperfine coupling, highlighting the importance of accounting for parameter correlations in multiparameter quantum metrology.

Having examined the precision of magnetic-field estimation, we now turn to the estimation of the hyperfine coupling strength \(\Lambda\). Figure \ref{figure3} displays the corresponding minimum variance as a function of the magnetic field and temperature. In Fig. \ref{figure3}(a), the uncertainty generally increases with \(\Lambda\), indicating a deterioration of the estimation precision as the hyperfine interaction becomes stronger. Its dependence on \(\mathcal B\), however, is not uniform throughout the parameter space. According to Eq. (\ref{eq:bound_Lambda}), the attainable precision is determined by the full QFIM, including the correlations between the simultaneously estimated parameters. This dependence reflects the fact that the magnetic field and hyperfine coupling jointly determine the energy spectrum, and consequently influence the sensitivity of the thermal state to variations in \(\Lambda\).

The temperature dependence shown in Fig. \ref{figure3}(b) exhibits a more structured behavior. In particular, a region of relatively large uncertainty emerges at low temperatures and strong hyperfine coupling, whereas lower values are found over an intermediate-temperature region. This nonmonotonic behavior indicates that thermal effects do not simply degrade the estimation precision. Rather, the redistribution of thermal populations among the energy eigenstates modifies the sensitivity of the Gibbs state to \(\Lambda\), leading to parameter-dependent changes in the QFIM and its inverse. The resulting landscape therefore identifies regions in which the hyperfine coupling can be estimated with comparatively higher precision. Together, these observations highlight the distinct but interconnected roles of the magnetic field and thermal fluctuations in determining the precision of hyperfine-coupling estimation.
	\begin{figure}[b]
	\centering
	\includegraphics[width=0.95\linewidth]{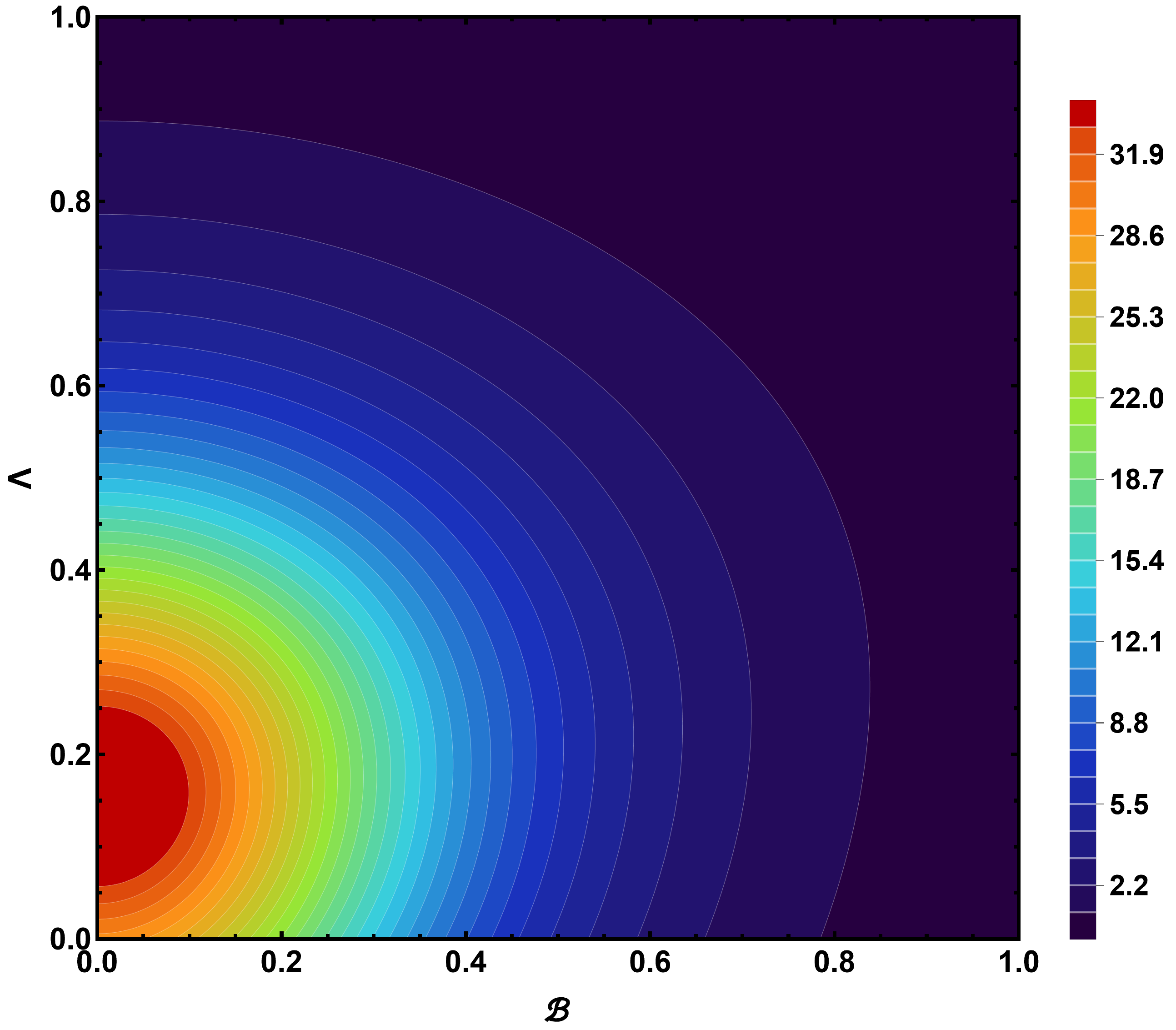}
\caption{Contour plot of the determinant of the QFIM, $\det(\mathcal N)$, as a function of $\mathcal B$ and $\Lambda$, for $T=0.2$, $\mu=1$, and $\nu=\mu/1836.14$.}
	\label{figure4}
\end{figure}
To further characterize the joint estimation of the magnetic field and hyperfine coupling, Fig. \ref{figure4} presents the determinant of the QFIM as a function of \(\mathcal B\) and \(\Lambda\). A pronounced maximum emerges in the weak-field regime at intermediate hyperfine coupling strengths, approximately \(\Lambda\simeq0.15\)–\(0.2\), whereas the determinant decreases substantially as either parameter increases. In multiparameter quantum estimation, the QFIM provides a matrix description of the information available about the unknown parameters and determines the corresponding quantum Cram\'er–Rao bounds \cite{Liu2020,Szczykulska2016}. Accordingly, its determinant characterizes the overall information content associated with the two-parameter estimation problem and indicates how well-conditioned the corresponding estimation problem is. The observed maximum therefore identifies a region where the two parameters are jointly distinguishable with greater information content. Conversely, the reduction of \(\det(\mathcal N)\) at stronger fields and couplings indicates a deterioration of the joint estimation precision. Nevertheless, this global indicator should be considered together with the individual variances and the normalized correlation coefficient, since a large determinant alone does not guarantee optimal precision for each parameter separately.
	\begin{figure}[b]
	\centering
	\includegraphics[width=0.95\linewidth]{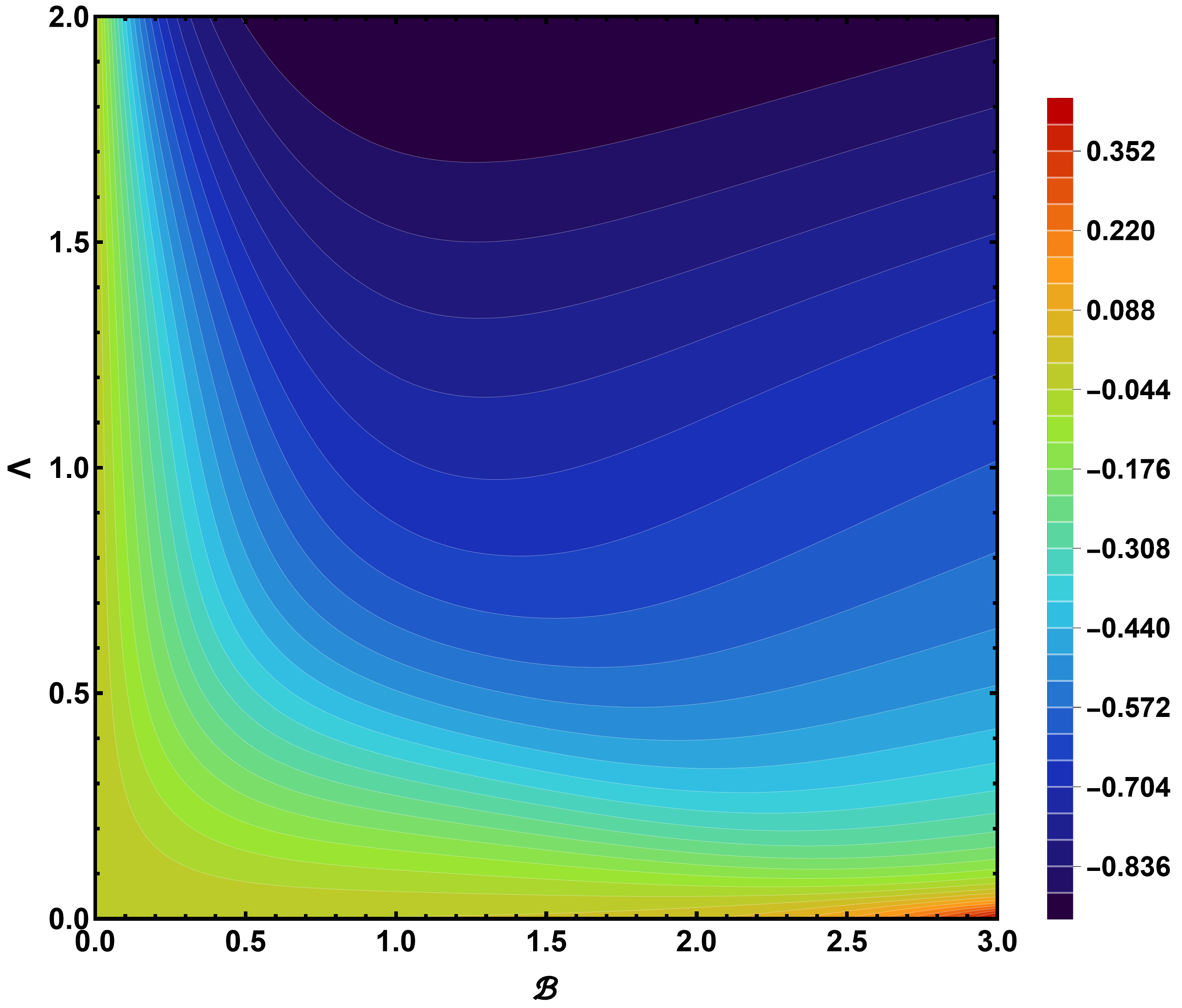}
\caption{Contour plot of the normalized metrological correlation coefficient $\mathcal R_{\mathcal B\Lambda}$ as a function of $\mathcal B$ and $\Lambda$, for $T=0.2$, $\mu=1$, and $\nu=\mu/1836.14$.}
	\label{figure5}
\end{figure}
The normalized metrological correlation coefficient \(\mathcal R_{\mathcal B\Lambda}\), displayed in Fig. \ref{figure5}, provides complementary information on the joint estimation of the magnetic field and hyperfine coupling. Unlike the determinant of the QFIM, which characterizes the overall information content, this coefficient quantifies the relative contribution of the off-diagonal information term to the two-parameter estimation problem. As shown in the figure, its sign and magnitude vary significantly across the parameter space. Positive values occur predominantly at weak hyperfine coupling, particularly toward the larger-field region, whereas negative values prevail over most of the remaining domain. This sign reversal indicates that the information coupling between the two parameters is strongly dependent on the operating point of the system.

Such behavior is consistent with the general structure of multiparameter quantum estimation, where the simultaneous estimation of different parameters is governed by both diagonal and off-diagonal elements of the QFIM \cite{Liu2020,Szczykulska2016}. In the present model, this dependence originates from the joint contribution of \(\mathcal B\) and \(\Lambda\) to the energy spectrum and to the thermal-state response. In particular, changes in either parameter modify the same energy splitting, thereby producing a parameter-dependent coupling between their respective sensitivities. The regions where \(|\mathcal R_{\mathcal B\Lambda}|\) approaches unity correspond to a strong relative coupling between the two information directions, whereas values closer to zero indicate a weaker coupling. Consequently, this figure complements the uncertainty and determinant analyses by identifying parameter regions in which the simultaneous estimation is more strongly affected by interparameter correlations.

Taken together, Figs. \ref{figure2}–\ref{figure5} provide a complementary characterization of the multiparameter estimation problem. While the individual variance bounds quantify the precision achievable for each parameter, the determinant of the QFIM describes the overall information content, and the normalized off-diagonal coefficient reveals the relative coupling between the two estimation directions. Their combined analysis shows that the metrological performance cannot be assessed through the sensitivity to either parameter independently. Instead, it results from the interplay between the magnetic field, the hyperfine interaction, and the thermal distribution of the system. This characterization establishes the metrological framework required to investigate, in the following section, how accurately the energetic properties of the same electron–proton system can be determined.
\section{Quantum Battery: Charging Dynamics and Energy-Storage Properties}
\label{sec:QB}

Having investigated the multiparameter estimation of the external
magnetic field $\mathcal B$ and the hyperfine interaction strength
$\Lambda$, we now turn to the quantum battery properties of the
electron--proton system. In this framework, the two-spin system is
regarded as a quantum battery whose initial state is the thermal
Gibbs state introduced in Sec.~\ref{sec:QFIM}. Our objective is to
characterize its charging dynamics and energy-storage capabilities
in terms of ergotropy, anti-ergotropy, storage capacity, and quantum
coherence. These quantities provide complementary information about
the energy stored in the system, the maximum extractable work, and
the quantum resources generated during the charging process
\cite{Allahverdyan2004,Binder2015,Campaioli2024}.

The charging process is implemented by applying an external
transverse driving field to the electron and proton spins. We
consider the charging Hamiltonian
\begin{equation}
	H_{\mathrm{ch}}
	=
	\omega
	\left(
	\sigma_x^{(e)}+\sigma_x^{(p)}
	\right),
	\label{eq:H_ch}
\end{equation}
where $\omega$ denotes the charging frequency and
$\sigma_x^{(e)}$ and $\sigma_x^{(p)}$ are the Pauli operators
acting on the electron and proton spins, respectively. Throughout
this work, we set $\hbar=1$. The corresponding unitary evolution
operator is
\begin{equation}
	U_{\mathrm{ch}}(t)
	=
	e^{-iH_{\mathrm{ch}}t}.
	\label{eq:U_ch}
\end{equation}

In the present work, we adopt an idealized controlled charging
protocol in which the intrinsic battery Hamiltonian $H$ is
temporarily switched off during the charging stage, while the
external transverse driving field described by
$H_{\mathrm{ch}}$ is applied. Consequently, the evolution during
this stage is governed solely by the charging Hamiltonian
$H_{\mathrm{ch}}$. At the end of the charging process, the
external driving field is switched off and the original
Hamiltonian $H$ is restored. The stored energy and the
corresponding work-related quantities are then evaluated with
respect to this Hamiltonian. The energetic costs associated
with switching the intrinsic Hamiltonian off and on are not
included in the present analysis. Thus, our results describe
the idealized charging performance of the system rather than
the total energetic cost of the complete protocol
\cite{Binder2015,Campaioli2024}.

For a constant charging field, the evolution operator can be
obtained analytically. In the computational basis
$\{|00\rangle,|01\rangle,|10\rangle,|11\rangle\}$, it reads
\begin{equation}
	U_{\mathrm{ch}}(t)
	=
	\begin{pmatrix}
		a(t)&c(t)&c(t)&b(t)\\
		c(t)&a(t)&b(t)&c(t)\\
		c(t)&b(t)&a(t)&c(t)\\
		b(t)&c(t)&c(t)&a(t)
	\end{pmatrix},
	\label{eq:U_ch_matrix}
\end{equation}
where
\begin{align}
	a(t)&=\cos^2(\omega t),\\
	b(t)&=-\sin^2(\omega t),\\
	c(t)&=-\frac{i}{2}\sin(2\omega t).
\end{align}

The state of the battery during the charging process is given by
the unitary transformation
\begin{equation}
	\rho(t)
	=
	U_{\mathrm{ch}}(t)	\rho_{\mathrm{th}}
	U_{\mathrm{ch}}^\dagger(t),
	\label{eq:rho_ch}
\end{equation}
where $	\rho_{\mathrm{th}}=e^{-H/T}/Z$ is the initial Gibbs state and
$Z=\operatorname{Tr}(e^{-H/T})$ is the partition function.
Since the charging dynamics is unitary, the eigenvalues of
$\rho(t)$ remain unchanged throughout the evolution. In
particular, the von Neumann entropy is conserved, whereas the
mean energy of the battery, evaluated with respect to the
intrinsic Hamiltonian $H$, can vary with time. The energy
stored during the charging process is therefore quantified by
\begin{equation}
	\Delta E(t)
	=
	\operatorname{Tr}[\rho(t)H]
	-
	\operatorname{Tr}[	\rho_{\mathrm{th}} H].
	\label{eq:stored_energy}
\end{equation}

However, the energy stored in a quantum battery does not
necessarily coincide with the amount of work that can be
extracted from it. The latter is quantified by the ergotropy,
which represents the maximum work extractable through cyclic
unitary transformations of the system
\cite{Allahverdyan2004,Binder2015}. For a general state
$\rho(t)$, the ergotropy is defined as
\begin{equation}
	\mathcal W_{\mathrm{QB}}(t)
	=
	\operatorname{Tr}[\rho(t)H]
	-
	\operatorname{Tr}[\rho_{\mathrm{pas}}(t)H],
	\label{eq:ergotropy}
\end{equation}
where $\rho_{\mathrm{pas}}(t)$ denotes the passive state
associated with $\rho(t)$. A passive state is a state from
which no work can be extracted by means of unitary
transformations. It is diagonal in the energy eigenbasis,
with its eigenvalues arranged in non-increasing order as
the corresponding energies increase.

Let $\{\varepsilon_j,|\varepsilon_j\rangle\}_{j=1}^{4}$ denote the eigenvalues and corresponding eigenstates of the Hamiltonian, rearranged in non-decreasing order according to
\begin{equation}
	\varepsilon_1\leq\varepsilon_2\leq\varepsilon_3\leq\varepsilon_4.
\end{equation}
These ordered eigenvalues are obtained by sorting the analytical energy levels $E_1,E_2,E_3,E_4$ given in Eqs.~(7) and (8), with the corresponding eigenstates relabeled accordingly. Let $\{p_j\}_{j=1}^{4}$ denote the eigenvalues of $\rho(t)$, arranged in non-increasing order:
\begin{equation}
	p_1\geq p_2\geq p_3\geq p_4.
\end{equation}
The passive state is then expressed as
\begin{equation}
	\rho_{\mathrm{pas}}(t)
	=
	\sum_{j=1}^{4}
	p_j|\varepsilon_j\rangle\langle \varepsilon_j|.
	\label{eq:passive_state}
\end{equation}

In our charging protocol, the initial Gibbs state is passive
with respect to the battery Hamiltonian. Moreover, since
the charging evolution is unitary, the eigenvalues of the
density operator are preserved. Consequently, the passive
state associated with $\rho(t)$ coincides with the initial
Gibbs state, namely
\begin{equation}
	\rho_{\mathrm{pas}}(t)=	\rho_{\mathrm{th}}.
	\label{eq:passive_gibbs}
\end{equation}
Substituting this relation into Eq.~(\ref{eq:ergotropy}),
we obtain
\begin{equation}
	\mathcal W_{\mathrm{QB}}(t)
	=
	\operatorname{Tr}
	\left[
	\big(\rho(t)-	\rho_{\mathrm{th}}\big)H
	\right].
	\label{eq:ergotropy_gibbs}
\end{equation}
It therefore follows from Eqs.~(\ref{eq:stored_energy})
and~(\ref{eq:ergotropy_gibbs}) that
\begin{equation}
		\mathcal W_{\mathrm{QB}}(t)=\Delta E(t).
	\label{eq:energy_ergotropy}
\end{equation}
For the present electron–proton system, substituting the explicit expressions of the thermal-state matrix elements into Eq. (\ref{eq:ergotropy_gibbs}) yields the following analytical expression for the ergotropy
\begin{equation}
		\mathcal{W}_{\mathrm{QB}}(t)
		=-2\sin^2(\omega t)\Big[
	\delta\left(\rho_{11}-\rho_{44}\right)\\
		+\Delta\left(\rho_{22}-\rho_{33}\right)
		\Big].
	\label{eq:explicit_ergotropy}
\end{equation}
Thus, within the considered protocol, the energy accumulated
during charging is entirely extractable as useful work.
This equality follows from the passivity of the initial Gibbs
state and the preservation of the density-matrix spectrum
under unitary evolution. It is specific to the present
charging protocol and does not hold for arbitrary initial
battery states.
\begin{figure*}[t]
	\centering
	\includegraphics[width=0.38\linewidth]{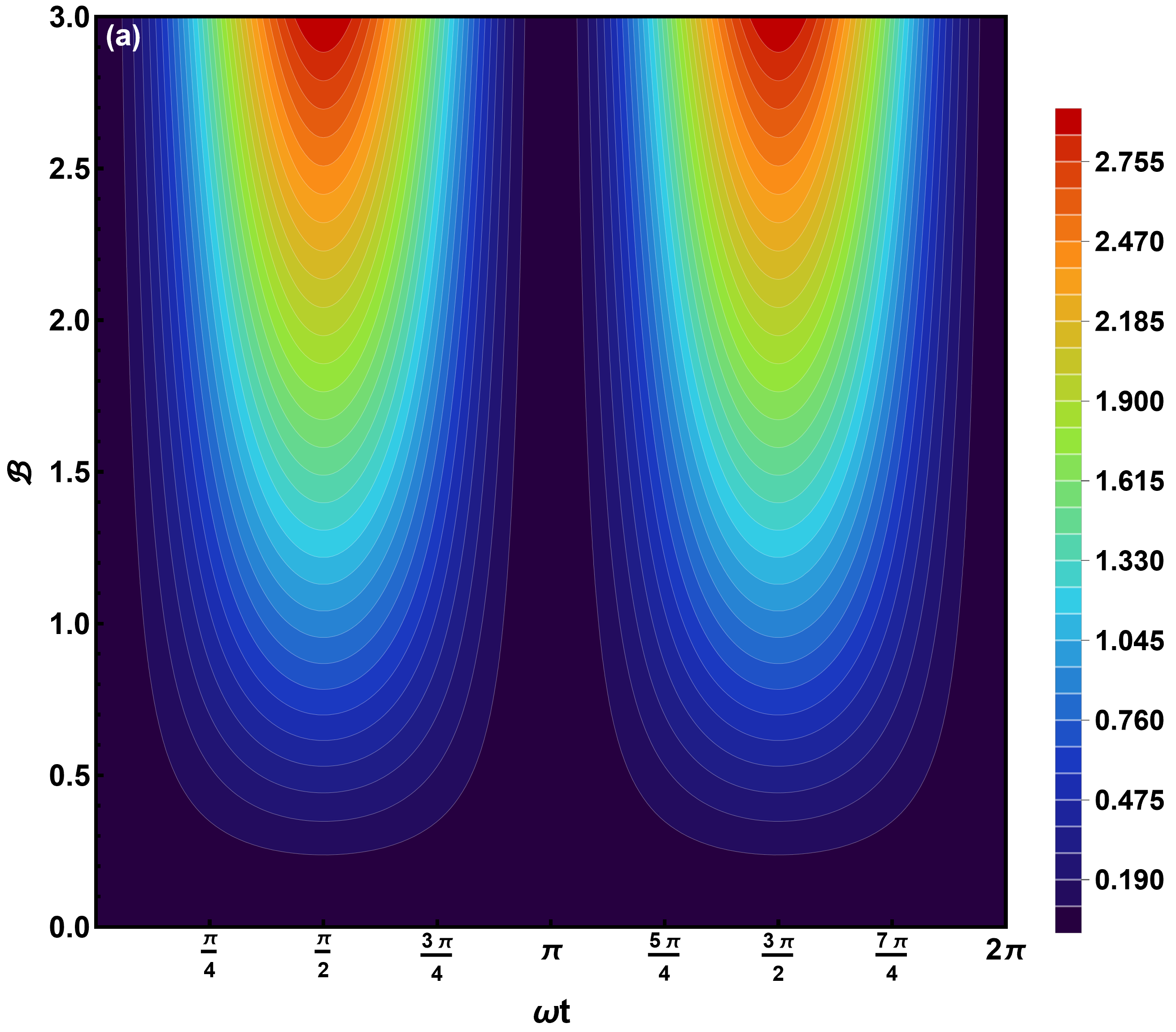}\hspace*{1.4cm}
	\includegraphics[width=0.38\linewidth]{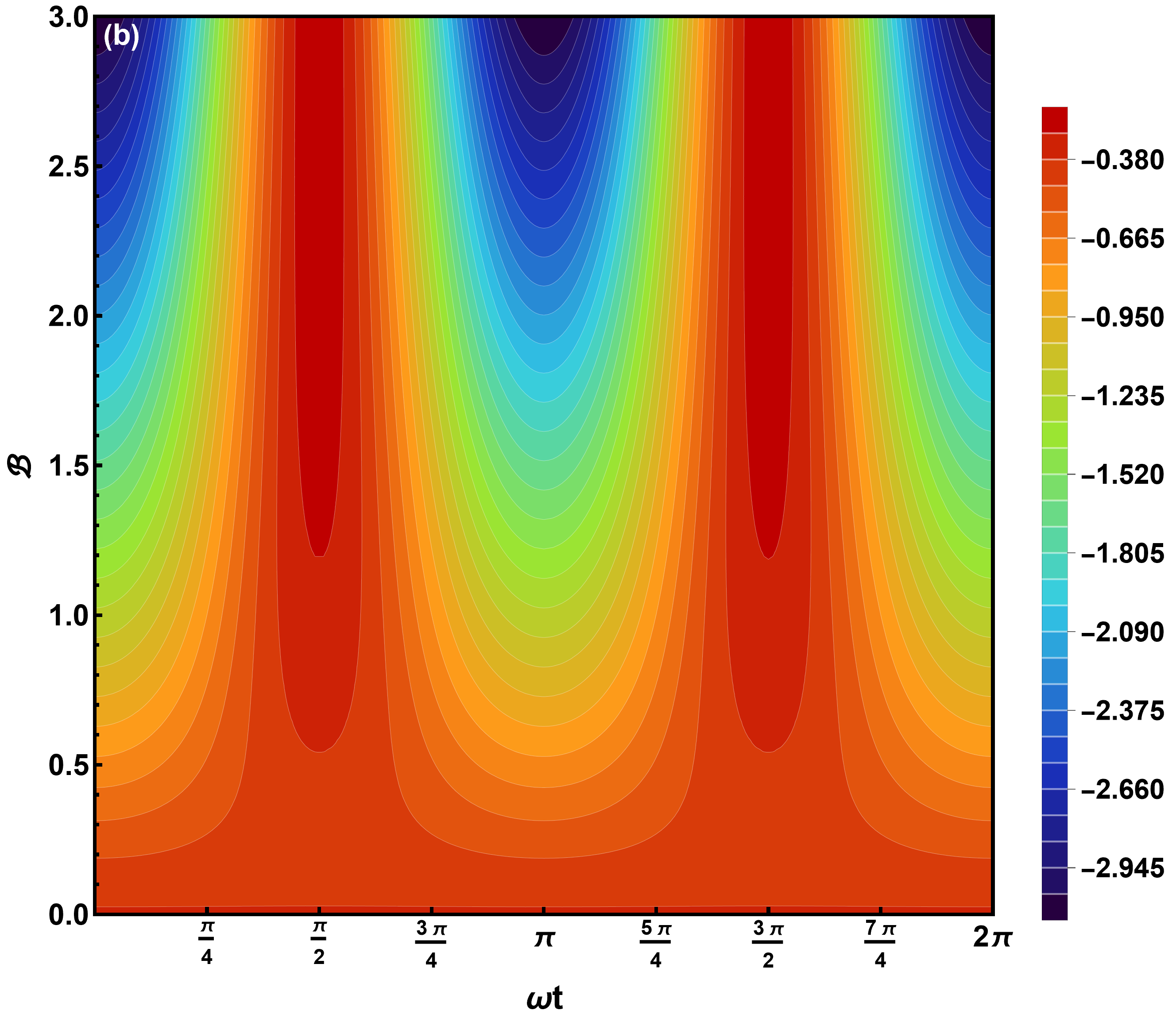}
	\includegraphics[width=0.38\linewidth]{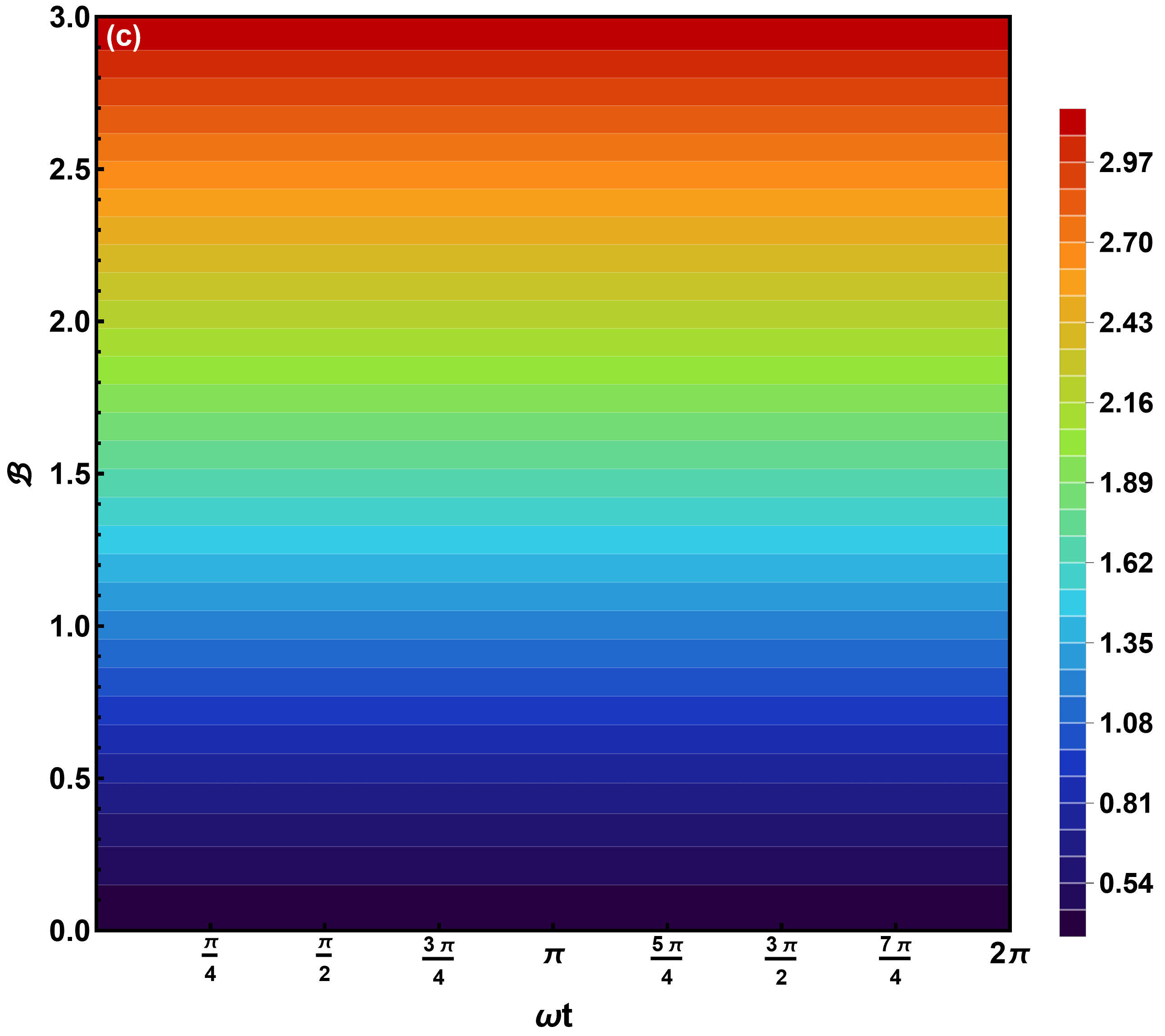}\hspace*{1.4cm}	\includegraphics[width=0.38\linewidth]{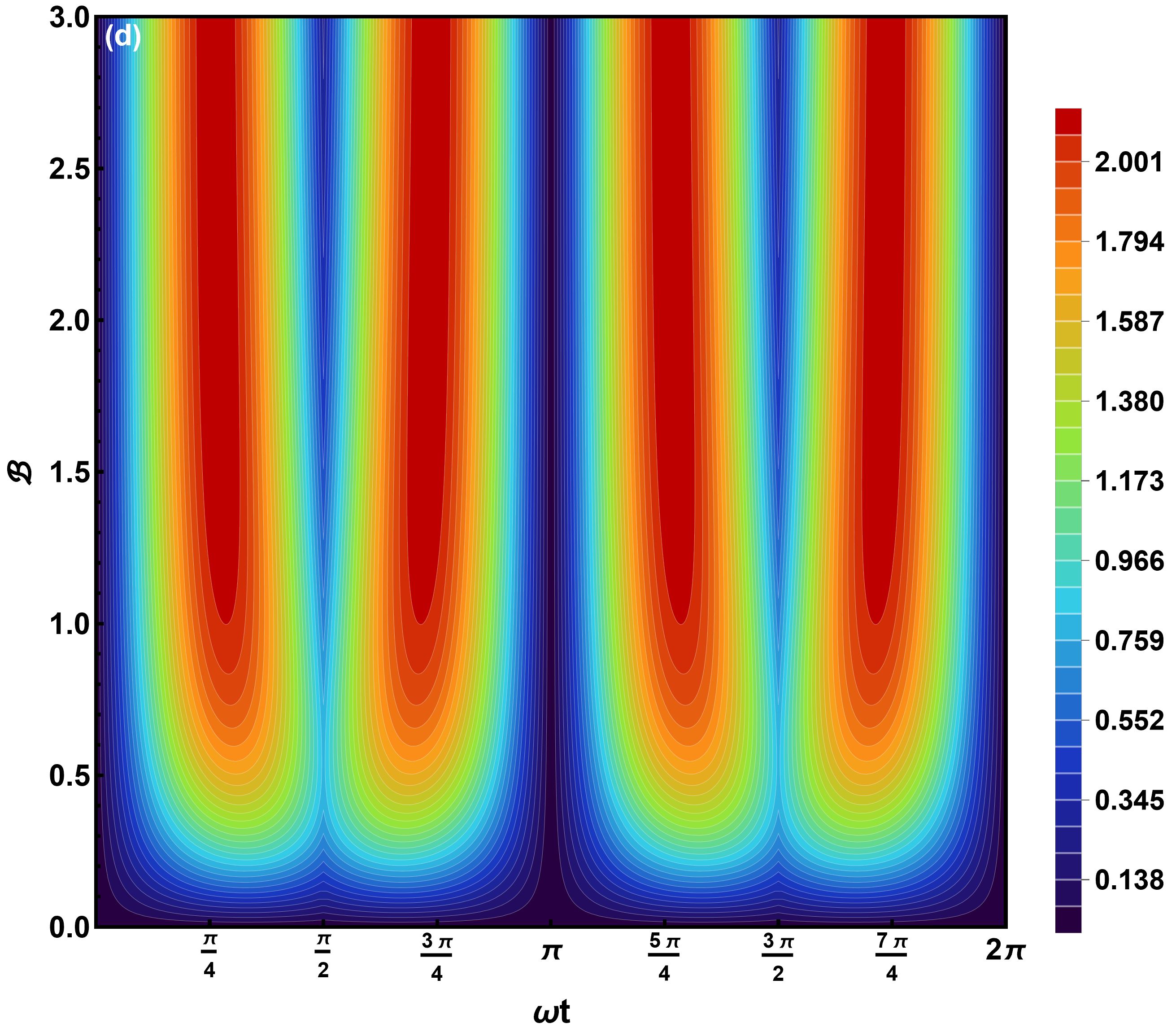}
	\caption{Contour plots of (a) ergotropy $\mathcal W_{\mathrm{QB}}$, (b) anti-ergotropy $\overline{\mathcal W}_{\mathrm{QB}}$, (c) storage capacity $\mathcal C_{\mathrm a}$, and (d) $\ell_1$-norm of quantum coherence $\mathcal C_{\ell_1}$ as functions of the dimensionless charging time $\omega t$ and the external magnetic field $\mathcal B$, for $\Lambda=0.5$, $T=0.2$, $\mu=1$, and $\nu=\mu/1836.14$.}
	\label{figure6}
\end{figure*}
\begin{figure*}
	\includegraphics[width=0.38\linewidth]{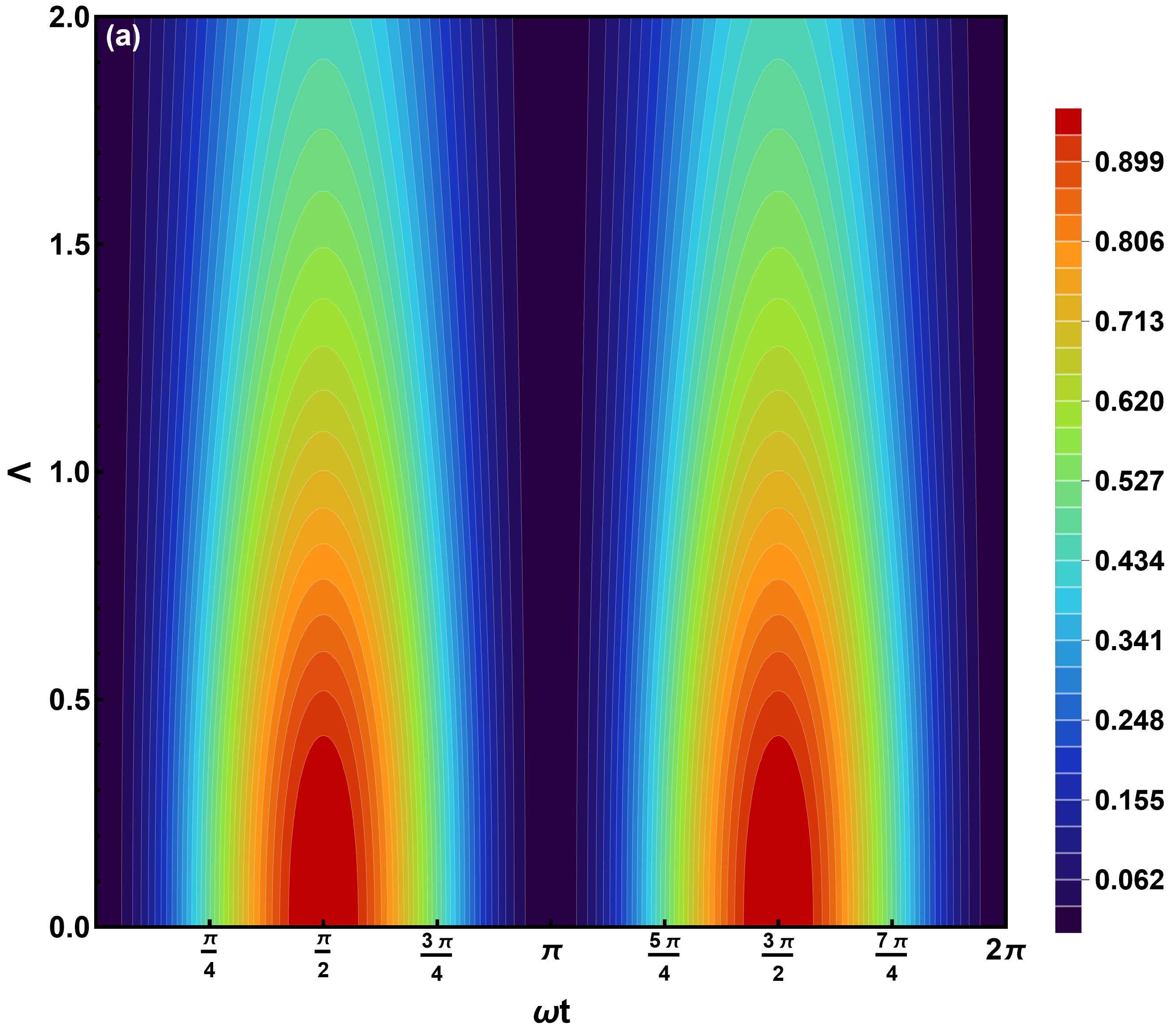}\hspace*{1.4cm}
	\includegraphics[width=0.38\linewidth]{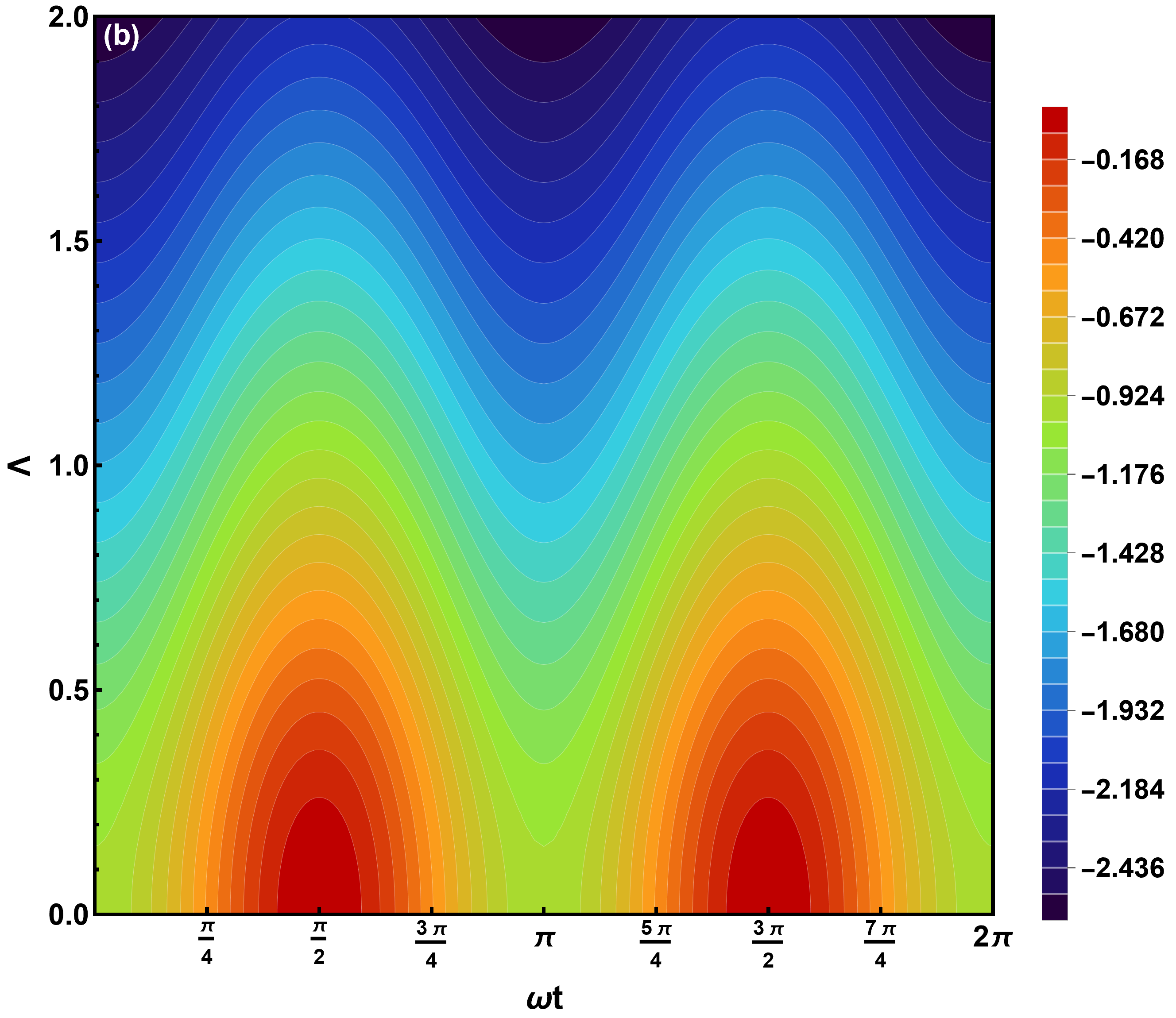}
	\includegraphics[width=0.38\linewidth]{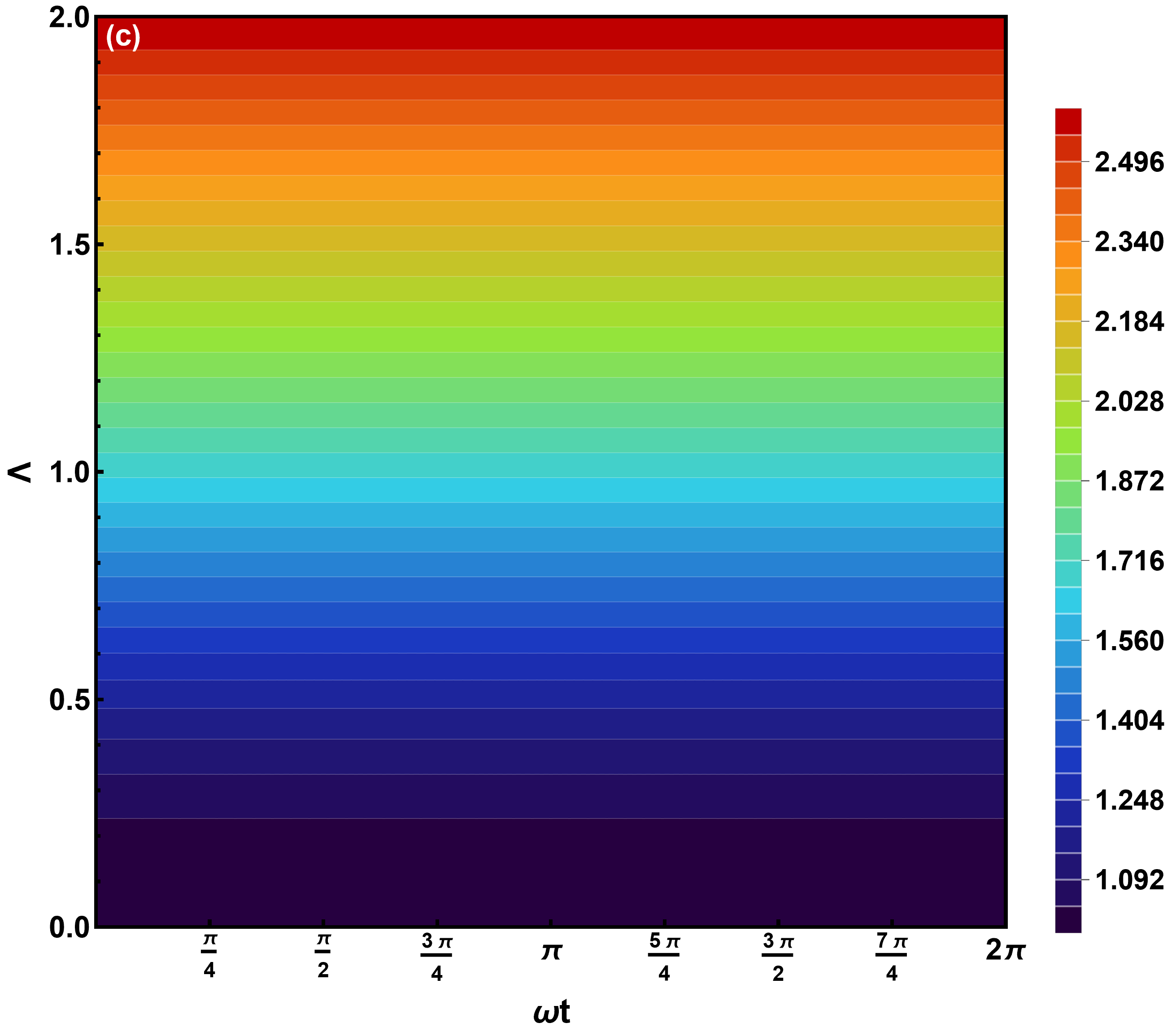}\hspace*{1.4cm}	\includegraphics[width=0.38\linewidth]{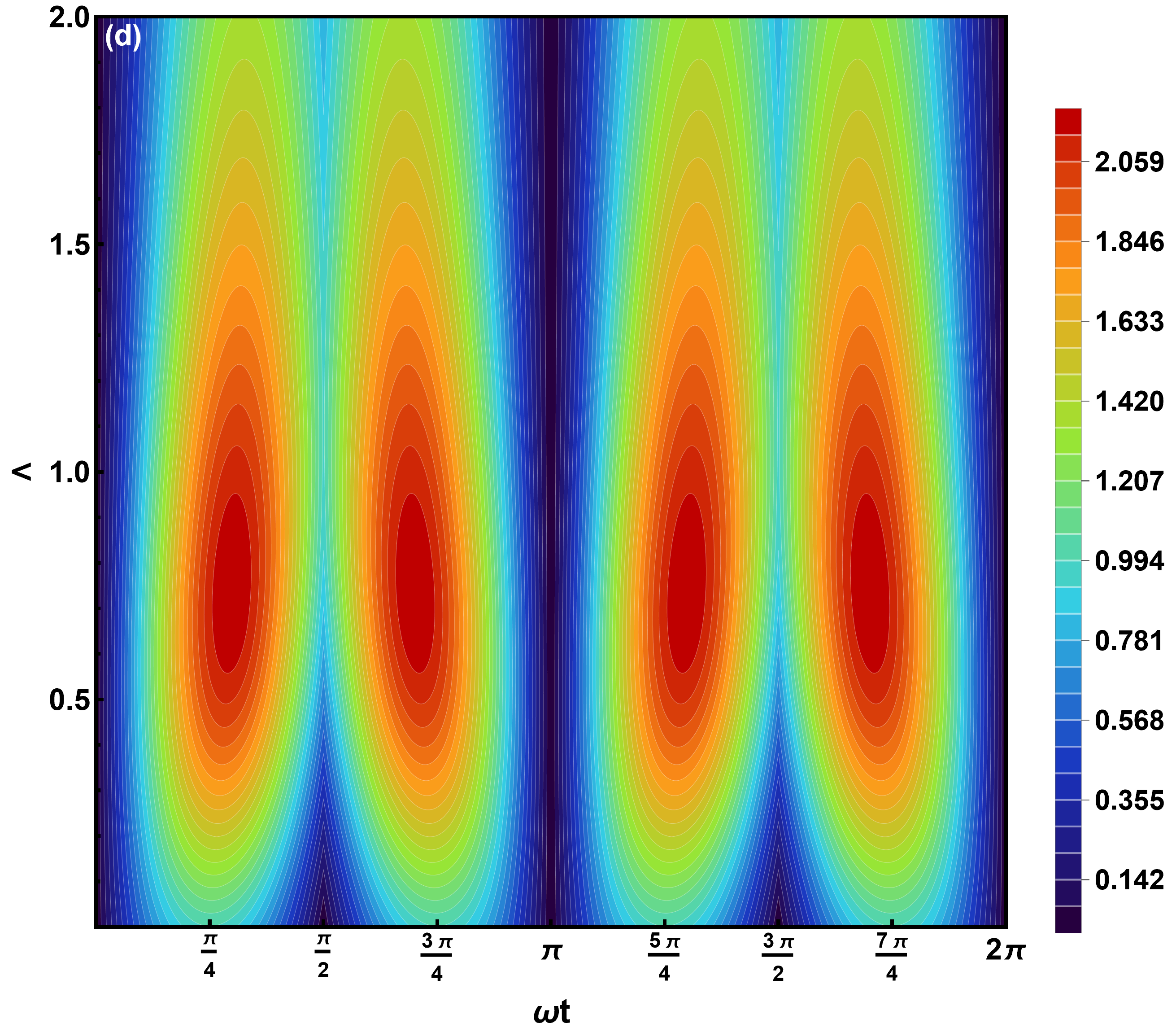}
	\caption{Contour plots of (a) ergotropy $\mathcal W_{\mathrm{QB}}$, (b) anti-ergotropy $\overline{\mathcal W}_{\mathrm{QB}}$, (c) storage capacity $\mathcal C_{\mathrm a}$, and (d) $\ell_1$-norm of quantum coherence $\mathcal C_{\ell_1}$ as functions of the dimensionless charging time $\omega t$ and the hyperfine interaction strength $\Lambda$, for $\mathcal B=1$, $T=0.2$, $\mu=1$, and $\nu=\mu/1836.14$.}
	\label{figure7}
\end{figure*}
\begin{figure*}
	\centering
	\includegraphics[width=0.38\linewidth]{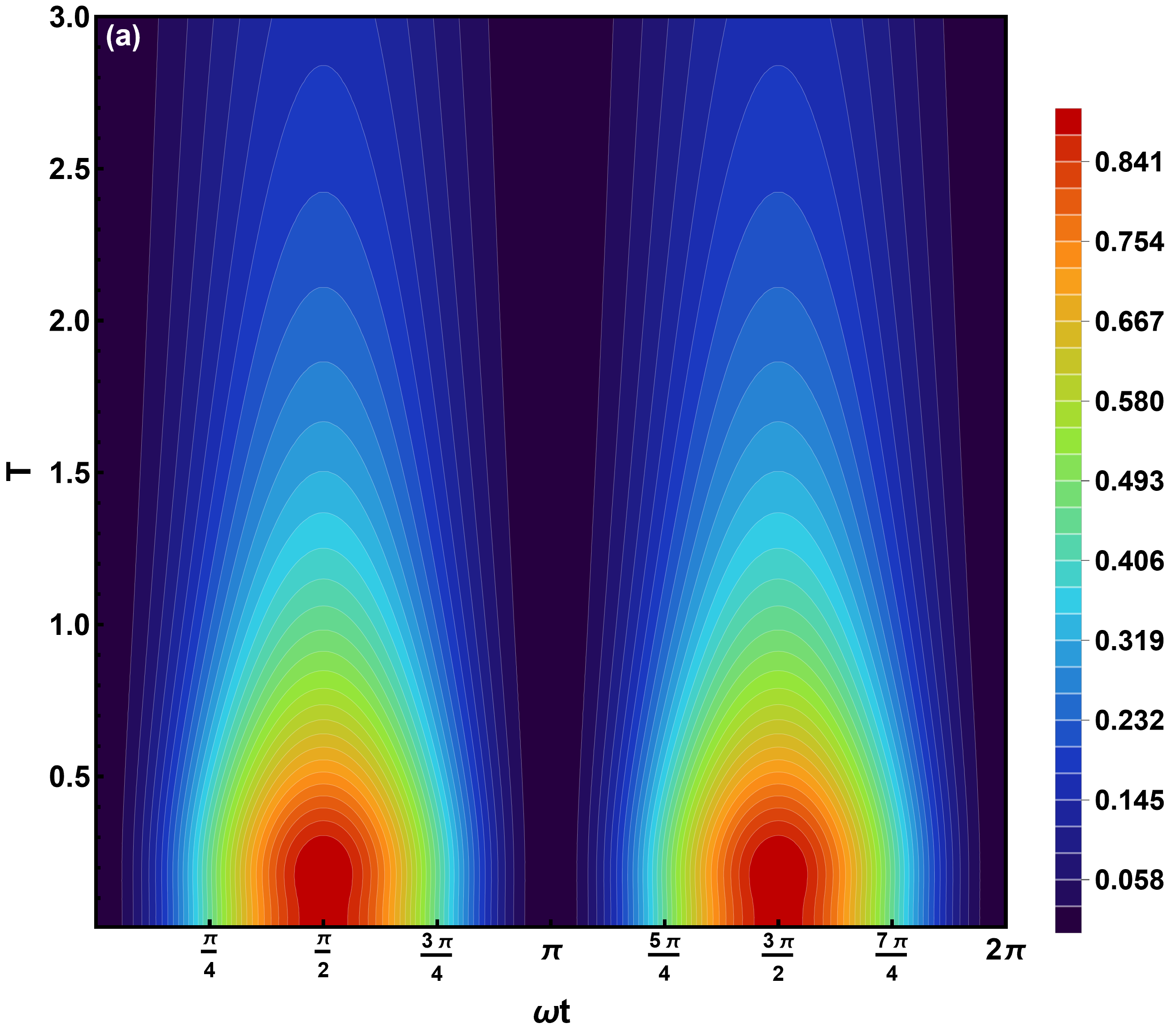}\hspace*{1.4cm}
	\includegraphics[width=0.38\linewidth]{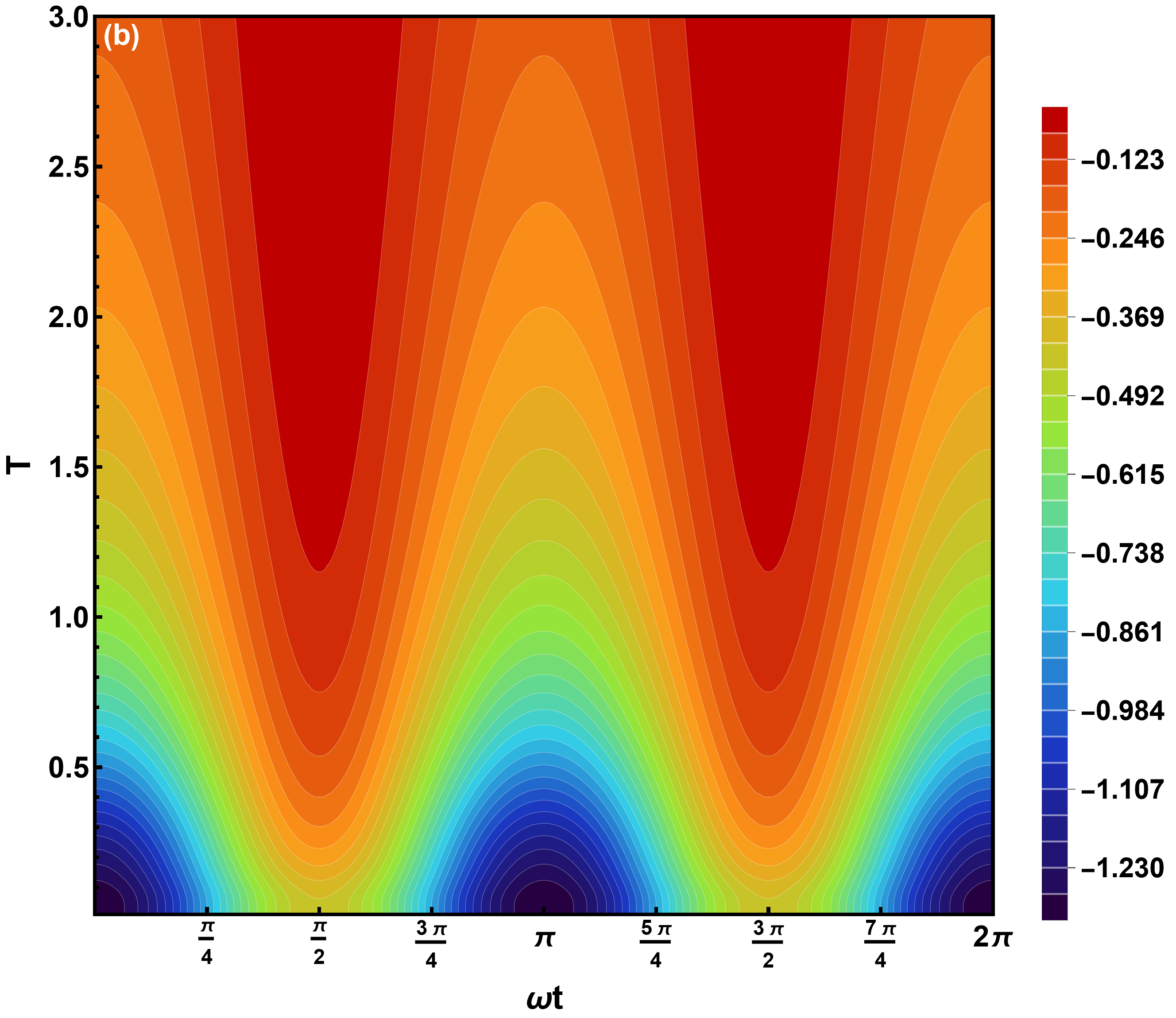}
	\includegraphics[width=0.38\linewidth]{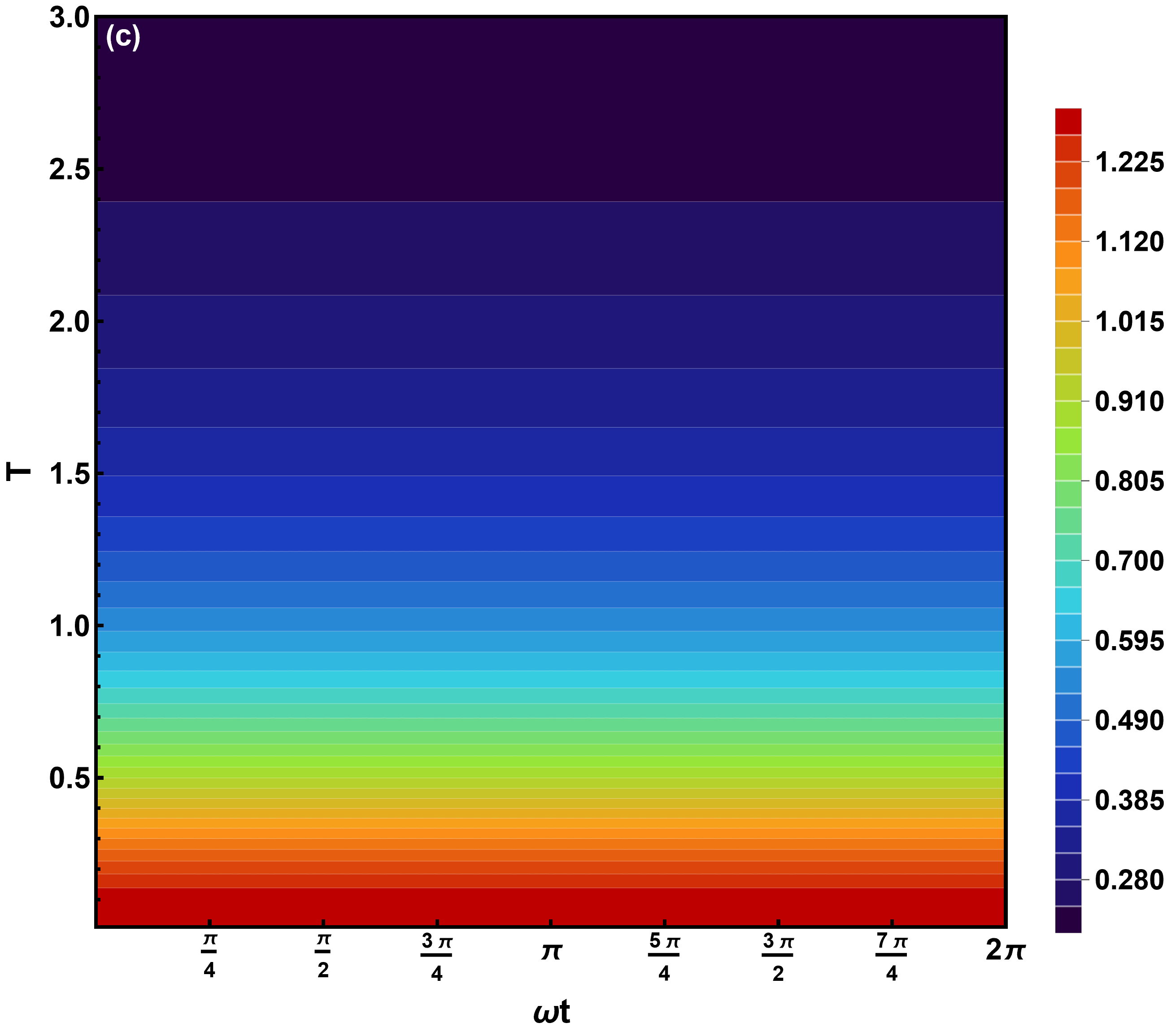}\hspace*{1.4cm}	\includegraphics[width=0.38\linewidth]{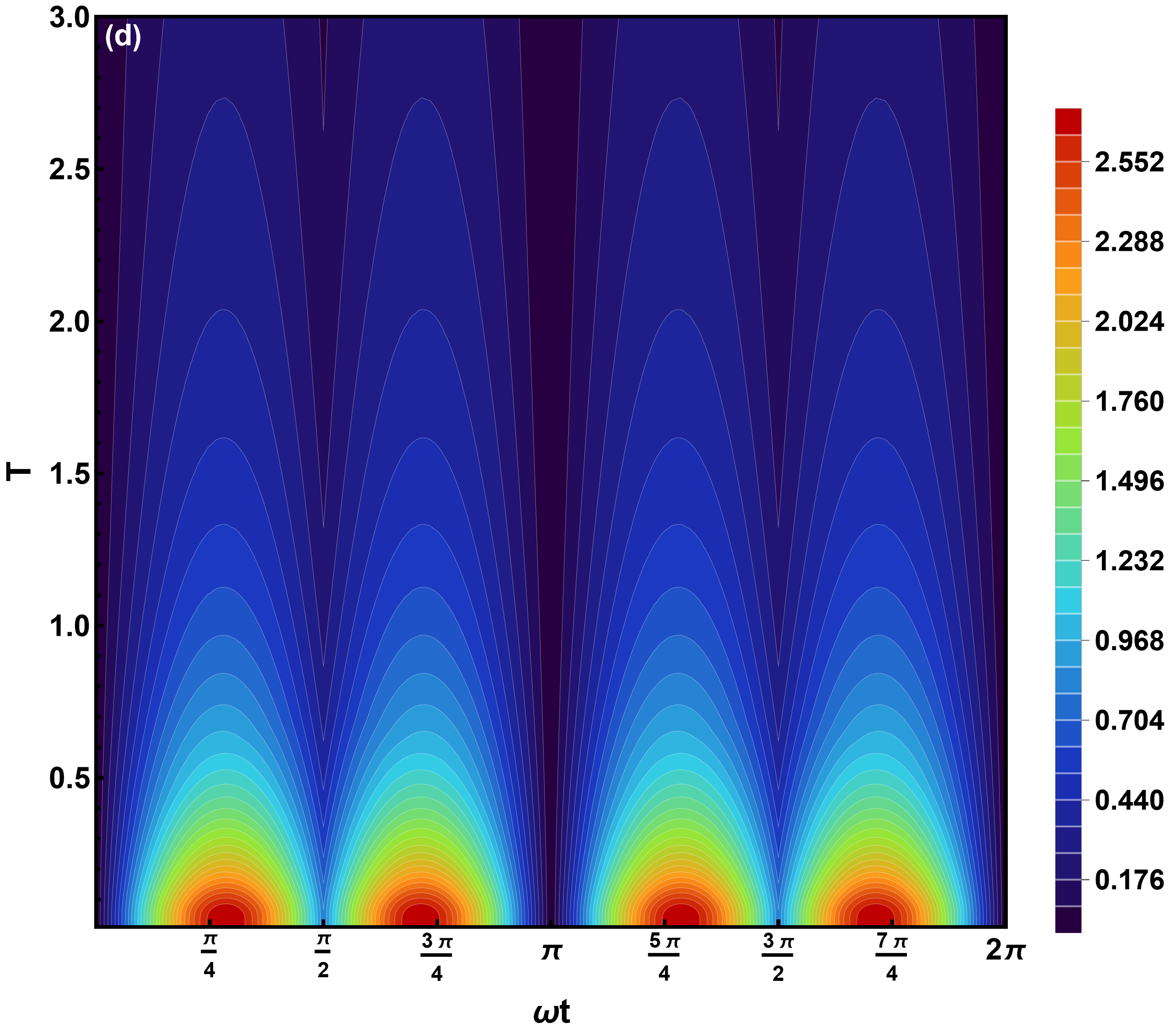}
	\caption{Contour plots of (a) ergotropy $\mathcal W_{\mathrm{QB}}$, (b) anti-ergotropy $\overline{\mathcal W}_{\mathrm{QB}}$, (c) storage capacity $\mathcal C_{\mathrm a}$, and (d) $\ell_1$-norm of quantum coherence $\mathcal C_{\ell_1}$ as functions of the dimensionless charging time $\omega t$ and the temperature \(T\), for $\mathcal B=1$, $\Lambda=0.5$, $\mu=1$, and $\nu=\mu/1836.14$.}
	\label{figure8}
\end{figure*}
To characterize the opposite energetic limit, we introduce
the anti-ergotropy, which describes the maximum amount of
energy that can be supplied to the battery through a cyclic
unitary transformation. This quantity is associated with
the active state, obtained by assigning the largest
eigenvalues of the density operator to the highest energy
eigenstates. For the present four-dimensional system, the
active state is given by
\begin{equation}
	\rho_{\mathrm{act}}(t)
	=
	\sum_{j=1}^{4}
	p_j|\varepsilon_{5-j}\rangle\langle \varepsilon_{5-j}|.
	\label{eq:active_state}
\end{equation}
Following the signed convention adopted here, the
anti-ergotropy is defined as
\begin{equation}
	\overline{\mathcal W}_{\mathrm{QB}}(t)
	=
	\operatorname{Tr}
	\left[
	\big(\rho(t)-\rho_{\mathrm{act}}(t)\big)H
	\right].
	\label{eq:anti_ergotropy}
\end{equation}
With this convention, the anti-ergotropy is non-positive.
Its absolute value represents the maximum energy that can
be injected into the battery by a cyclic unitary
transformation. The active state corresponds to the
maximum-energy state accessible within the unitary orbit
of $\rho(t)$.

The storage capacity of a quantum battery quantifies the
energy range accessible through unitary transformations.
It is defined as the difference between the ergotropy and
the signed anti-ergotropy
\cite{Yang2023},
\begin{equation}
	\mathcal C_{\mathrm a}
	=
	\mathcal W_{\mathrm{QB}}(t)
	-
	\overline{\mathcal W}_{\mathrm{QB}}(t).
	\label{eq:storage_capacity}
\end{equation}
Equivalently, it can be written as
\begin{equation}
	\mathcal C_{\mathrm a}
	=
	\operatorname{Tr}
	\left[
	\big(\rho_{\mathrm{act}}(t)
	-\rho_{\mathrm{pas}}(t)\big)H
	\right].
	\label{eq:capacity_energy_range}
\end{equation}
Since the eigenvalues of $\rho(t)$ are conserved during
unitary charging, both the passive and active states are
time-independent. Therefore, the storage capacity is
constant throughout the charging process:
\begin{equation}
	\frac{d\mathcal C_{\mathrm a}}{dt}=0.
	\label{eq:capacity_invariance}
\end{equation}
This invariance reflects the fact that the capacity is an
intrinsic property determined by the spectrum of the
battery state and its Hamiltonian, rather than by the
instantaneous charging stage. In contrast, the ergotropy
describes the extractable work actually accumulated during
the charging process.

Finally, to investigate the role of quantum coherence in
the charging dynamics, we consider the $\ell_1$-norm of
coherence. Coherence is a fundamental quantum resource
that quantifies the contribution of off-diagonal
matrix elements in a specified reference basis
\cite{Baumgratz2014}. In the present work, we choose the
energy eigenbasis $\{|E_j\rangle\}_{j=1}^{4}$ of the
battery Hamiltonian as the reference basis. The
$\ell_1$-norm of coherence is then defined as
\begin{equation}
	\mathcal C_{\ell_1}[\rho(t)]
	=
	\sum_{\substack{i,j=1\\i\neq j}}^{4}
	\left|
	\langle E_i|\rho(t)|E_j\rangle
	\right|.
	\label{eq:l1_coherence}
\end{equation}

This quantity measures the total magnitude of the
off-diagonal elements of the charged state in the energy
eigenbasis. Since the initial Gibbs state is diagonal in
this basis, any nonzero value of
$\mathcal C_{\ell_1}[\rho(t)]$ reflects the coherence
generated by the charging operation. Its time dependence
therefore provides insight into the coherent features
accompanying the energy-storage process.

In the following, we investigate the dependence of the
ergotropy $\mathcal W_{\mathrm{QB}}(t)$, the
anti-ergotropy $\overline{\mathcal W}_{\mathrm{QB}}(t)$,
the storage capacity $\mathcal C_{\mathrm a}$, and the
$\ell_1$-norm of coherence
$\mathcal C_{\ell_1}[\rho(t)]$ on the charging time and
the physical parameters of the electron--proton system,
namely the external magnetic field $\mathcal B$, the
hyperfine interaction strength $\Lambda$, and the
temperature $T$. This analysis allows us to explore how
the magnetic field, spin--spin interaction, and thermal
initialization jointly affect the energy-storage
properties and quantum coherence of the battery.

Having established the theoretical framework and defined the relevant energetic and coherence quantifiers, we now turn to the numerical investigation of the charging performance of the electron–proton quantum battery. In particular, we examine how the external magnetic field \(B\), the hyperfine coupling strength \(\Lambda\), and the initial temperature \(T\) influence the charging dynamics and the associated quantum resources. The results are presented through contour plots, which allow us to identify the parameter regions associated with enhanced work extraction, storage capacity, and coherence generation. We also investigate the relationship between these quantities by comparing the original charged state with its energy-dephased counterpart. This analysis provides a basis for understanding the role of the physical parameters in determining the energetic performance of the battery.

Figure \ref{figure6} provides a comparative picture of the energetic and coherence-related properties of the quantum battery as functions of the dimensionless charging time \(\omega t\) and the external magnetic field \(\mathcal B\). As shown in Fig. \ref{figure6}(a), the ergotropy exhibits a periodic behavior, with pronounced maxima around \(\omega t=\pi/2\) and \(3\pi/2\). This oscillatory structure follows directly from the \(\sin^2(\omega t)\) dependence of the stored extractable energy obtained from the charging dynamics. Moreover, the ergotropy generally increases with the magnetic field, indicating that the external field provides an additional means of controlling the amount of extractable energy.

The anti-ergotropy displayed in Fig. \ref{figure6}(b) exhibits a complementary behavior, with its most negative values occurring around \(\omega t=\pi\). This evolution reflects the changes in the energetic ordering of the charged state relative to its associated active state, according to the sign convention adopted here. In contrast, Fig. \ref{figure6}(c) reveals that the storage capacity remains independent of the charging time, although it varies with \(\mathcal B\). This time independence follows from the unitary nature of the charging protocol, which preserves the eigenvalue spectrum of the density operator and, consequently, the capacity defined in terms of this spectrum.

Finally, Fig. \ref{figure6}(d) shows that the \(\ell_1\)-norm of coherence undergoes periodic oscillations, with maxima occurring at approximately the same charging times as those of the ergotropy. This temporal correspondence suggests that coherence generation accompanies the charging process and may contribute to the formation of extractable energy. Nevertheless, the similarity between their oscillatory patterns does not by itself establish a direct proportionality between coherence and ergotropy. Overall, these results reveal distinct dynamical roles of energy storage and quantum coherence: while the extractable work and coherence evolve periodically during charging, the storage capacity remains invariant under the considered unitary dynamics.

The dependence of the battery properties on the hyperfine interaction strength is further examined in Fig. \ref{figure7}, with the external magnetic field fixed at \(\mathcal B=1\). Figure \ref{figure7}(a) shows that the ergotropy retains its periodic dependence on the charging time, with maxima around \(\omega t=\pi/2\) and \(3\pi/2\). However, their magnitude decreases as \(\Lambda\) increases, indicating that a stronger hyperfine interaction does not necessarily enhance work extraction. This behavior highlights the role of the internal electron–proton coupling in determining the amount of energy that can be extracted from the charged state.

A complementary feature is observed for the anti-ergotropy in Fig. \ref{figure7}(b), whose most negative values occur around \(\omega t=\pi\), particularly at larger coupling strengths. This dependence reflects the modification of the energy distribution associated with the active state as the hyperfine interaction is varied. In contrast, Fig. \ref{figure7}(c) exhibits a monotonic increase in the storage capacity with \(\Lambda\), while remaining independent of the charging time. This behavior follows from the invariance of the density-matrix spectrum under unitary charging and indicates that the coupling strength can enhance the storage capacity without producing a corresponding increase in the instantaneous extractable work.

The coherence profile in Fig. \ref{figure7}(d) displays a different dependence on the hyperfine interaction. In particular, its maxima occur within an intermediate range of \(\Lambda\), revealing a nonmonotonic response to the coupling strength. This feature suggests that the generation of coherence is optimized under specific interaction conditions rather than increasing continuously with \(\Lambda\). The comparison between Figs. \ref{figure7}(a), \ref{figure7}(c), and \ref{figure7}(d) therefore reveals an important distinction: the coupling strength that favors storage capacity is not necessarily the one that maximizes either ergotropy or coherence. Consequently, the optimal operating conditions of the battery depend on which energetic or quantum resource is considered.

The thermal influence on the battery performance is investigated in Fig. \ref{figure8}, where the external magnetic field and hyperfine coupling are fixed at \(\mathcal B=1\) and \(\Lambda=0.5\), respectively. Figure \ref{figure8}(a) shows that the ergotropy exhibits a pronounced periodic dependence on the charging time, with maxima around \(\omega t=\pi/2\) and \(3\pi/2\). More importantly, these maxima decrease as the temperature increases, indicating that thermal mixing progressively limits the amount of extractable work. This behavior highlights the detrimental role of thermal fluctuations in the charging performance of the considered system.

A related thermal dependence is observed for the anti-ergotropy in Fig. \ref{figure8}(b), whose values become less negative with increasing temperature. In contrast, the storage capacity shown in Fig. \ref{figure8}(c) remains independent of the charging time, as expected from the invariance of the density-matrix spectrum under unitary evolution. Its decrease with temperature indicates that thermal mixing also reduces the energy-storage potential of the battery, independently of the charging stage. Finally, Fig. \ref{figure8}(d) reveals that the coherence exhibits a periodic structure with four maxima over the interval \(0\leq\omega t\leq2\pi\). The amplitude of these maxima decreases with temperature, showing that thermal fluctuations suppress the coherence generated during the charging process. Taken together, these results establish that increasing temperature degrades both the extractable work and the storage capacity, while simultaneously reducing the coherence generated by the charging dynamics. Thus, low-temperature operation is favorable for preserving the energetic and coherence-related resources of the quantum battery.

\section{Conclusion}\label{sec:cc}

In this work, we have investigated the multiparameter quantum metrology and energetic properties of an electron--proton spin system inspired by the hyperfine structure of the hydrogen atom. By considering the external magnetic field $\mathcal B$ and the hyperfine coupling strength $\Lambda$ as unknown parameters, we derived the quantum Fisher information matrix and the corresponding quantum Cram\'er--Rao bounds. Our results reveal a nontrivial dependence of the estimation precision on the magnetic field, hyperfine interaction, and temperature. Furthermore, although the associated symmetric logarithmic derivatives do not commute in general, the weak-commutativity condition is satisfied, supporting the asymptotic compatibility of the simultaneous estimation problem.

We also analyzed the energetic properties of the system as a quantum battery, focusing on ergotropy, anti-ergotropy, storage capacity, and energy-basis coherence. The ergotropy exhibits periodic charging dynamics, whereas the storage capacity remains constant under the considered unitary evolution. The dependence of these quantities on the magnetic field, hyperfine coupling, and temperature highlights that the optimal conditions for work extraction, energy storage, and coherence generation do not necessarily coincide. Moreover, comparison with the energy-dephased state shows that energy-basis coherence enhances the extractable work and is accompanied by an increased storage capacity in the investigated regime.

Overall, our findings provide a comprehensive characterization of the metrological and energetic properties of a hyperfine-coupled spin system. They highlight its potential relevance for multiparameter quantum sensing and quantum energy storage, while illustrating the distinct roles played by the external magnetic field, hyperfine interaction, and thermal effects in these two quantum applications.

\appendix
\section{Explicit expressions of the QFIM elements}
\label{app:QFIM}

For completeness, we provide here the explicit analytical
expressions of the QFIM elements introduced in
Sec.~\ref{sec:QFIM}. The diagonal and off-diagonal elements
are respectively given by
	\begin{widetext}
		\begin{align}\nonumber
			\mathcal N_{\mathcal B\mathcal B}
			={}&
			\frac{1}{
				\rho_{11}(\rho_{22}+\rho_{33})
				(\rho_{22}\rho_{33}-\rho_{23}^{2})
			}
			\Bigg\{
			-(\partial_{\mathcal B}\rho_{11})^2
			\rho_{22}\rho_{23}^{2}
		+2(\partial_{\mathcal B}\rho_{33})
			\rho_{11}\rho_{23}
			\left[
			-2(\partial_{\mathcal B}\rho_{23})\rho_{22}
			+(\partial_{\mathcal B}\rho_{22})\rho_{23}
			\right]
			\\\nonumber
			&+4(\partial_{\mathcal B}\rho_{23})^2
			\rho_{11}\rho_{22}\rho_{33}
			+(\partial_{\mathcal B}\rho_{11})^2
			\rho_{22}^{2}\rho_{33}
		-4(\partial_{\mathcal B}\rho_{22})
			(\partial_{\mathcal B}\rho_{23})
			\rho_{11}\rho_{23}\rho_{33}
			-(\partial_{\mathcal B}\rho_{11})^2
			\rho_{23}^{2}\rho_{33}
			\\\nonumber
			&+(\partial_{\mathcal B}\rho_{11})^2
			\rho_{22}\rho_{33}^{2}
			+(\partial_{\mathcal B}\rho_{33})^2
			\rho_{11}
			\left[
			-\rho_{23}^{2}
			+\rho_{22}(\rho_{22}+\rho_{33})
			\right]
		+(\partial_{\mathcal B}\rho_{22})^2
			\rho_{11}
			\left[
			-\rho_{23}^{2}
			+\rho_{33}(\rho_{22}+\rho_{33})
			\right]
			\Bigg\}
			\\
			&+\frac{(\partial_{\mathcal B}\rho_{44})^2}{\rho_{44}}.
	\\
			\mathcal N_{\mathcal B\Lambda}
			={}&\nonumber
			\frac{1}{
				\rho_{11}(\rho_{22}+\rho_{33})
				(\rho_{22}\rho_{33}-\rho_{23}^{2})
			}
			\Bigg\{
		-\rho_{23}\Bigg[
			2(\partial_{\mathcal B}\rho_{23})
			(\partial_{\Lambda}\rho_{33})
			\rho_{11}\rho_{22}
			+(\partial_{\mathcal B}\rho_{22})
			\left(
			\partial_{\Lambda}\rho_{22}
			-\partial_{\Lambda}\rho_{33}
			\right)
			\rho_{11}\rho_{23}
			\\\nonumber
			&\qquad
			+(\partial_{\mathcal B}\rho_{11})
			(\partial_{\Lambda}\rho_{11})
			\rho_{22}\rho_{23}
			\Bigg]
		+
			\Bigg[
			\rho_{22}\Big(
			(\partial_{\mathcal B}\rho_{22})
			(\partial_{\Lambda}\rho_{22})\rho_{11}
			+4(\partial_{\mathcal B}\rho_{23})
			(\partial_{\Lambda}\rho_{23})\rho_{11}
			+(\partial_{\mathcal B}\rho_{11})
			(\partial_{\Lambda}\rho_{11})\rho_{22}
			\Big)
			\\\nonumber
			&\qquad
			-2\Big(
			(\partial_{\mathcal B}\rho_{23})
			(\partial_{\Lambda}\rho_{22})
			+
			(\partial_{\mathcal B}\rho_{22})
			(\partial_{\Lambda}\rho_{23})
			\Big)
			\rho_{11}\rho_{23}
			-(\partial_{\mathcal B}\rho_{11})
			(\partial_{\Lambda}\rho_{11})\rho_{23}^{2}
			\Bigg]\rho_{33}
			\\\nonumber
			&\qquad+
			\Big[
			(\partial_{\mathcal B}\rho_{22})
			(\partial_{\Lambda}\rho_{22})\rho_{11}
			+
			(\partial_{\mathcal B}\rho_{11})
			(\partial_{\Lambda}\rho_{11})\rho_{22}
			\Big]\rho_{33}^{2}
		+
			(\partial_{\mathcal B}\rho_{33})\rho_{11}
			\Bigg[
			-(\partial_{\Lambda}\rho_{33})\rho_{23}^{2}
			\\
			&\qquad
			+\rho_{23}\Big(
			-2(\partial_{\Lambda}\rho_{23})\rho_{22}
			+(\partial_{\Lambda}\rho_{22})\rho_{23}
			\Big)
			+(\partial_{\Lambda}\rho_{33})
			\rho_{22}(\rho_{22}+\rho_{33})
			\Bigg]
			\Bigg\}
	+
			\frac{
				(\partial_{\Lambda}\rho_{44})
				(\partial_{\mathcal B}\rho_{44})
			}{\rho_{44}}
\\\nonumber
				\mathcal N_{\Lambda\Lambda}
				={}&
				\frac{1}{
					\rho_{11}(\rho_{22}+\rho_{33})
					(\rho_{22}\rho_{33}-\rho_{23}^{2})
				}
				\Bigg\{
				-(\partial_{\Lambda}\rho_{11})^2
				\rho_{22}\rho_{23}^{2}
		+2(\partial_{\Lambda}\rho_{33})\rho_{11}\rho_{23}
				\left[
				-2(\partial_{\Lambda}\rho_{23})\rho_{22}
				+(\partial_{\Lambda}\rho_{22})\rho_{23}
				\right]
				\\\nonumber
				&+4(\partial_{\Lambda}\rho_{23})^2
				\rho_{11}\rho_{22}\rho_{33}
				+(\partial_{\Lambda}\rho_{11})^2
				\rho_{22}^{2}\rho_{33}
				-4(\partial_{\Lambda}\rho_{22})
				(\partial_{\Lambda}\rho_{23})
				\rho_{11}\rho_{23}\rho_{33}
				-(\partial_{\Lambda}\rho_{11})^2
				\rho_{23}^{2}\rho_{33}
				\\\nonumber
				&+(\partial_{\Lambda}\rho_{11})^2
				\rho_{22}\rho_{33}^{2}
				+(\partial_{\Lambda}\rho_{33})^2
				\rho_{11}
				\left[
				-\rho_{23}^{2}
				+\rho_{22}(\rho_{22}+\rho_{33})
				\right]
			+(\partial_{\Lambda}\rho_{22})^2
				\rho_{11}
				\left[
				-\rho_{23}^{2}
				+\rho_{33}(\rho_{22}+\rho_{33})
				\right]
				\Bigg\}
				\\
				&+\frac{(\partial_{\Lambda}\rho_{44})^2}{\rho_{44}}.
			\end{align}
	\end{widetext}
These expressions are obtained by substituting the analytical
SLD operators into the defining relations of the QFIM.

\end{document}